\documentclass[%
 reprint,
 amsmath,amssymb,
 aps,
]{revtex4-2}

\usepackage{graphicx}% Include figure files
\usepackage{subfigure}
\usepackage{microtype}
\usepackage{booktabs}
\usepackage{dcolumn}% Align table columns on decimal point
\usepackage{bm}% bold math
\usepackage{tikz}
\usepackage{calligra}
\usepackage{amsmath}
\usepackage{ifpdf}
\ifpdf
\usepackage{hyperref}
\fi
\begin{document}

\preprint{APS/123-QED}
\title{Community Detection via Katz and Eigenvector Centrality}
%\thanks{Code available at \textellipsis}%

\author{Mark Ditsworth}
 \email{markditsworth@protonmail.com}
% \noaffiliation
\author{Justin Ruths}%
 \email{jruths@utdallas.edu}
 \affiliation{University of Texas at Dallas, Richardson, Texas 75080}

\date{\today}% It is always \today, today,

\begin{abstract}
The computational demands of community detection algorithms such as Louvain and spectral optimization can be prohibitive for large networks.
Eigenvector centrality and Katz centrality are two network statistics commonly used to describe the relative importance of nodes; and their calculation can be closely approximated on large networks by scalable iterative methods.
In this paper, we present and leverage a surprising relationship between Katz centrality and eigenvector centrality to detect communities. Beyond the computational gains, we demonstrate that our approach identifies communities that are as good or better than conventional methods.
\end{abstract}

%Use showkeys class option if keyword
\maketitle

%\tableofcontents

\section{Introduction}
\label{intro}
    In complex network analysis, community detection is often employed to find meaningful insights into the underlying organization of nodes. 
	Discerning which nodes are best grouped together provides great utility for applications such as efficient data routing \cite{DTN}, electronic virus and worm containment \cite{worm1}, as well as aiding in network reconstruction \cite{reconstruct}. 
	%Most popular community detection algorithms rely on iterative heuristics to attempt modularity maximization. The combinatoric nature of these algorithms make their runtime ungainly for large networks, thus requiring expensive computing power to achieve results in a reasonable amount of time.
	Most community detection algorithms assign node class such that modularity is maximized. 
	Modularity (Q) is a measure of the isolation between communities, and is calculated by the difference between the observed fraction of edges between like nodes (i.e., nodes of the same class) and the fraction of edges expected at random \cite{Mod}, 
	\begin{equation}
		\label{eqn:mod2}
		Q = \frac{1}{2m}\sum_{i=1}^{n}\sum_{j=1}^{n}\left(A_{ij} -\frac{k_ik_j}{2m}\right)\delta(c_i,c_j),
	\end{equation}
    for a graph with $n$ nodes and $m$ edges, $n\times n$ adjacency matrix $A$, and where $k_i$ and $c_i$ are the degree and class of node $i$.
    
	Thus, for $Q>0$ there are more edges between like nodes than would be expected at random, indicating assortative mixing (segregation between node classes). Likewise, for $Q<0$ there are fewer edges between like nodes than expected, indicating disassortative mixing.
	
	While in theory $Q$ is bounded above by 1, most network structures do not result in 0 expected edges between like nodes with randomly assigned edges. Therefore, a network with $Q$ greater than 0 signifies assortative mixing, but not necessarily the \textit{strength} of assortative mixing. For that we must find the maximum possible value of $Q$, $Q_{max}$,
	
	\begin{equation}
	\label{eqn:mod_max}
	Q_{max} = 1 - \frac{1}{2m}\sum_{i=1}^{n}\sum_{j=1}^{n}\frac{k_ik_j}{2m}~\delta(c_i,c_j).
	\end{equation}
	The second term is the same as that of (\ref{eqn:mod2}), but all edges are assumed to be between like nodes by setting the first term to 1. Normalizing $Q$ by $Q_{max}$ gives a more complete quantifier of the strength of assortative mixing. $Q$ may only be 0.3, but if $Q_{max}$ is only 0.34, then the network is nearly as assortatively mixed as it can be. Thus it is common to show normalized modularity, or to show $Q_{max}$ in addition to $Q$.
	
	True maximization of modularity is NP-complete \cite{NP}, so modularity-maximization algorithms implement iterative heuristics to guide the search.
	Simple modularity maximization \cite{simpleMM2} and the Louvain method \cite{louvain} are both based on the process of iterating through node class assignments, and selecting them based on maximum increase (or minimal decrease) in the modularity of the communities. In comparison, spectral modularity maximization \cite{spectral} attempts to maximize modularity analytically using a relaxation of the original mixed-integer optimization problem, ultimately by finding the leading eigenvector of the defined modularity matrix.
	
	Each of these community detection algorithms requires either numerous iterations through combinatorial partitions of the network nodes or linear algebraic operations on the adjacency matrix. More recent literature on community detection extends the use of these methods, altering calculations and processes, but do not deviate from the iterative maximization of modularity \cite{Leider,alpha}. For large complex networks the computational and memory requirements often prove impractical.
	
	This work demonstrates the utility of Katz centrality and eigenvector centrality as indicators of community membership in large undirected networks. This method is shown to produce well-defined communities when sufficient modularity is present in the network, which is a limitation shared by all modularity-based community detection methods. The proposed method is also shown to complete in a much faster runtime. Based on our datasets our proposed approach has runtimes as low as 8.6\% of the Louvain community detection runtime for smaller networks, and 0.02\% of the Louvain runtime for larger networks.
	
	Section \ref{s:cent} details the mathematical principles behind the proposed community detection method. Section \ref{s:AggCluster} describes the clustering algorithm used to label community members. Section \ref{s:randomGraphs} demonstrates the behavior of the Katz vs. eigenvector centrality plots using ad-hoc modular networks formed from random graphs. Section \ref{s:realWorld} illustrates the utility of eigenvector and Katz centrality as features for community detection in real-world networks. Finally, Section \ref{s:qual} provides compares the proposed community detection method to the widely-used Louvain method. All associated code supporting this work is made available \cite{github}.

\section{Eigenvector and Katz Centrality}
	\label{s:cent}
	Eigenvector centrality is based on the idea that a node's importance is related to the importance of its neighbors.
	The eigenvector centrality of node $i$ ($x_i$) is measured by the scaled sum of the eigenvector centralities of its neighbors, 
	\begin{equation}
	\label{eqn:evc_m}
	\mathbf{x} = \frac{1}{\lambda_1}\mathbf{A}\mathbf{x},
	\end{equation}
	where $\lambda_1$ is the leading eigenvalue of the adjacency matrix $\mathbf{A}$, and $\mathbf{x} = [x_1, x_2, \dots, x_n]^T$ \cite{EVC}.
	
	This is clearly the eigenvector equation; hence the name eigenvector centrality. For large networks, calculating the spectral decomposition of $\mathbf{A}$ is computationally demanding.
	But (\ref{eqn:evc_m}) can be solved iteratively, until $\mathbf{x}^{(k)} \approx \mathbf{x}^{(k+1)}$ by
	\begin{equation}
	\label{eqn:evc}
	\mathbf{x}^{(k+1)} = \frac{1}{\lambda_1}\mathbf{A}~\mathbf{x}^{(k)}.
	\end{equation}
	This results in a close approximation of the leading eigenvector of $\mathbf{A}$ without consuming as much memory as directly solving (\ref{eqn:evc_m}).
	
	Katz centrality is calculated similarly to eigenvector centrality, but with free centrality $\beta$ given to all nodes \cite{katz},
	\begin{equation}
	\label{eqn:katz}
	%x_i^{(k+1)} = \alpha\sum_{j=1}^{n}A_{ij}~x_j^{(k)} + \beta.
	\mathbf{x}^{(k+1)} = \alpha\mathbf{A}\mathbf{x}^{(k)} + \beta\mathbf{1}.
	\end{equation}
	Katz centrality is then,
	\begin{equation}
	\label{eqn:katz_m}
	\mathbf{x} = (I-\alpha \mathbf{A})^{-1}~\beta\textbf{1}.
	\end{equation}
	If $\alpha$ is 0, all nodes' centralities are $\beta$. If $\alpha$ is $\frac{1}{\lambda_1}$, $(I-\alpha A)$ is singular, the inverse is undefined and the centralities diverge. For $0 \leq \alpha < \frac{1}{\lambda_1}$, the centralities converge. This centrality measure was created to account for the fact that acyclic structures in directed graphs can yield nodes with zero eigenvector centrality. The free centrality guarantees all nodes have at least some centrality.
	
	It is evident from (\ref{eqn:evc}) and (\ref{eqn:katz}) that generally speaking, a node's Katz centrality is greater than it's eigenvector centrality. 
	%Normally, the exact value of a node's centrality is not so important as its relative value, so as to discern the order of importance. But the amount by which Katz centrality is greater than eigenvector centrality can serve as in important indicator, as well will see, for differences in local connectivity.
	However, the addition of free centrality has greater consequences. Recall that eigenvector centrality comes from the leading eigenvector of the adjacency matrix, thereby only describing the most dominant mode of the network. This causes the localization of eigenvector centrality commonly seen in modular networks \cite{localization}. Here localization is meant to express that the centrality is - to a greater extent than we desire - confined (localized) to a certain collection of nodes, despite the existence of other collections that appear to have similar importance.
	
	It is also argued in \cite{localization} that Katz centrality is a more robust measure of centrality than eigenvector centrality. From (\ref{eqn:katz_m}), the inverse operation can be expressed as a power series
	\begin{equation}
	\label{eqn:pwrseries}
	    (I - \alpha \mathbf{A})^{-1} = I + \alpha \mathbf{A} + \alpha^2 \mathbf{A}^2 + \alpha^3 \mathbf{A}^3 + \dotsi~.
	\end{equation}
	Since we consider undirected networks, the adjacency matrix $\mathbf{A}$ is symmetric and will the have orthogonal eigenvectors $\mathbf{u}_1,\mathbf{u}_2,\dots,\mathbf{u}_n$ with eigenvalues $\lambda_1,\lambda_2,\dots,\lambda_n$. We can represent $\beta \textbf{1}$ in this new eigenbasis as $a_1 \mathbf{u}_1 + a_2 \mathbf{u}_2 + \dotsi + a_n \mathbf{u}_n$, for some $a_1, \dots, a_n$. Combining with (\ref{eqn:pwrseries}), Katz centrality from (\ref{eqn:katz_m}) becomes
	\begin{equation}
	\label{eqn:new_katz}
	    \mathbf{x} = (I + \alpha \mathbf{A} + \alpha^2 \mathbf{A}^2 + \dotsi ) (a_1\mathbf{u}_1 + a_2\mathbf{u_2} + \dotsi + a_n\mathbf{u}_n) ~,\\
	\end{equation}
	which is equivalently expressed as
	\begin{equation}
	    \mathbf{x} = a_1\mathbf{u}_1\sum_{k=0}^{\infty}(\alpha\lambda_1)^k +
	    %a_2\mathbf{u}_2\sum_{k=0}^{\infty}(\alpha\lambda_2)^k + 
	    \dotsi +
	    a_n\mathbf{u}_n\sum_{k=0}^{\infty}(\alpha\lambda_n)^k.
	\end{equation}
	With the condition that $\alpha < \frac{1}{\lambda_1}$, each infinite sum will converge. Thus, it is clear that Katz centrality spans the entire eigenbasis of $\mathbf{A}$. The localization of centrality in modular networks will be significantly reduced with Katz centrality compared to eigenvector centrality. It is this presence or lack of localization in eigenvector and Katz centrality that when taken together makes them a reliable indicator of modularity in undirected networks.
	
	In fact, in this work we leverage the localization of eigenvector centrality against the robustness of Katz centrality in sufficiently modular networks to identify the communities that give rise to the observed modularity. In particular, if the Katz centralities of the nodes of a modular network were to be plotted against their eigenvector centralities (hereafter referred to as a KE plot), we expect that eigenvector centrality values of nodes in separate communities to be more distinct than they are using Katz centality.
	%If the Katz centralities of the nodes of a modular network were to be plotted against their eigenvector centralities (hereafter referred to as a KE plot), \textcolor{red}{we could expect the localization of eigenvector centrality to restrict the space occupied by one community in the eigenvector centrality dimension more so than the other}. However, both communities should be free to cover equal amount of range in the Katz centrality dimension. 
	Indeed, this is the picture we see in Figures \ref{fig:example_ba} and \ref{fig:example_er}.
	
\section{Clustering Algorithm}
\label{s:AggCluster}
    Unsupervised learning algorithms segment data into distinct clusters based on their spatial similarity rather than on prior knowledge of class membership. K-means \cite{kmeans}, DBSCAN \cite{dbscan}, and agglomerative clustering \cite{agglom} are commonly-used algorithms that identify clusters in unlabeled data. 
    The central focus of this article is to advance the idea of using both Katz and eigenvector centrality for community detection, so we have chosen a simple technique to provide clustering, using a line detection method similar to the Radon transformation \cite{radon}. 
    We expect future work to be able to optimize the clustering algorithm further.
    
\begin{figure}
        \centering
        \subfigure[]{\label{fig:line}
        \includegraphics[width=0.96\linewidth]{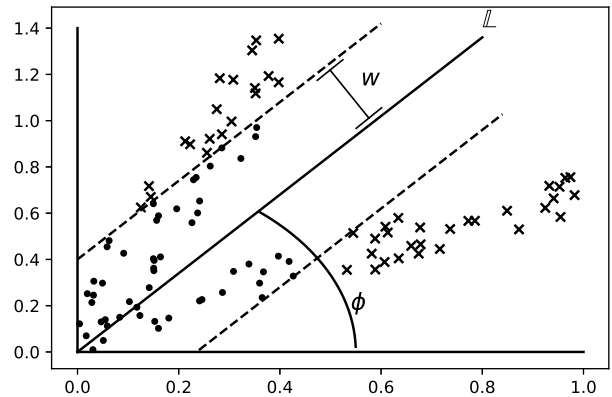}}
        \subfigure[]{\label{fig:s-theta}
        \includegraphics[trim={0 0 0 0.65cm},clip,width=0.96\linewidth]{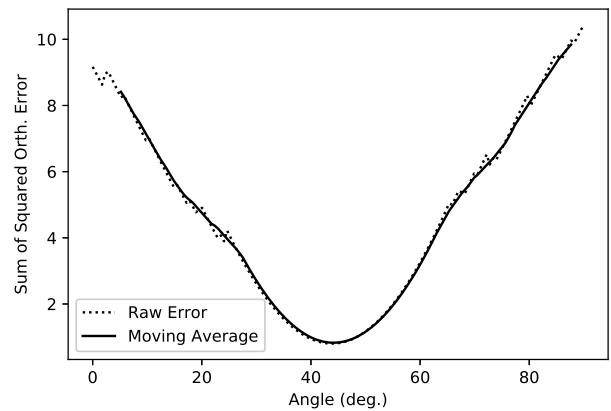}}
        \caption{\subref{fig:line} Example of line $\mathbb{L}(\phi_i)$ dividing the data points into two clusters, considering only data within $w$ orthogonal distance from $\mathbb{L}$. The exclusion of data with the $w$ parameter is necessary to allow for the detection of more than two communities. \subref{fig:s-theta} Example of finding optimal $\phi$ via identification of local minima. Here, $\phi=48^\text{o}$ is the only minimum, indicated two clusters in the data.}
    \end{figure}

    As seen in Figures \ref{fig:example_ba} and \ref{fig:example_er}, the clusters we seek to identify are aligned to distinct linear trends. Thus, we look to find lines that optimally divide the cluster regions, as depicted in Figure \ref{fig:line}. Line $\mathbb{L}(\phi)$ goes through the origin, forming angle $\phi$ with the $x$-axis. We then exclude data points that have greater than $w$ orthogonal distance from line $\mathbb{L}$, and calculate $\mathcal{S}$, the sum of squared orthogonal distances of the remaining points from line $\mathbb{L}$,
    \begin{equation}
        \label{eqn:cost_fcn}
        \mathcal{S} = \sum_i \left(x_i - \frac{y_i+x_i}{M^2+1}\right)^2 + \left(y_i - \frac{M^2y_i + Mx_i}{M^2 + 1}\right)^2~,
    \end{equation}
    where
    \begin{equation}
        M = \tan(\phi).
    \end{equation}
    Since the KE plot contains points in the first quadrant only, we sweep $\phi$ from 0$^o$ to 90$^o$, and look for local minima in the $\mathcal{S}$ vs. $\phi$ plot (after applying a moving average filter). For each $\phi_i$ that corresponds to a local minimum, we use $\mathbb{L}(\phi_i)$ as a decision boundary between clusters.
    
    This clustering method depends on three hyperparameters: $d$, the step size of the $\phi$ sweep, and the window size of the moving average filter. A primary benefit of this clustering method is the ability to detect multiple clusters, since it identifies $k+1$ clusters for $k$ local minima in the the $\mathcal{S}$ vs. $\phi$ plot.
    
\section{KE plots of Ad-hoc Modular Networks}
	\label{s:randomGraphs}
	We demonstrate the utility of the KE plot first on synthetic, ad-hoc modular networks. These networks are constructed to guarantee the existence of two communities, but with tunable modularity and relative density.
	
	\begin{figure*}
	    \centering
	    \subfigure[An example of the KE plot of one such modular graph with $n_1=n_2=250$ and $\mu=800$.]{
	    \label{fig:example_ba}
	    \includegraphics[width=0.46\linewidth]{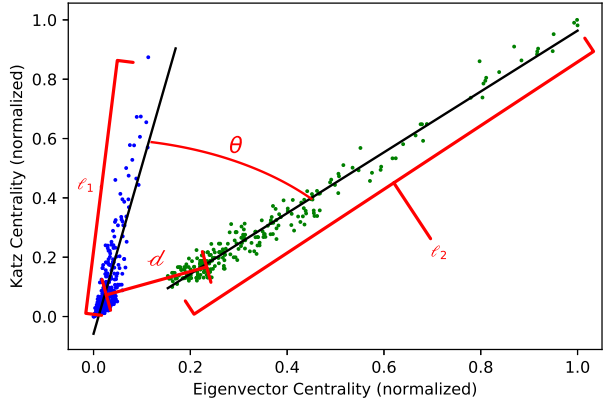}}\qquad
	    \subfigure[Plot of the average angle $\theta$ between the two clusters in the KE plot over 30 ad-hoc modular networks.]{
	    \label{fig:angle_ba}
	    \includegraphics[width=0.46\linewidth]{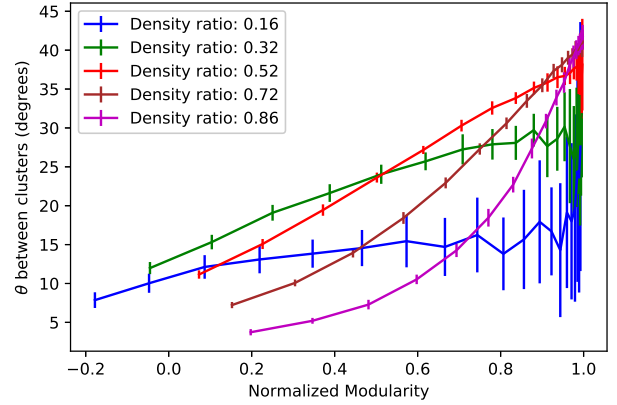}}\\
	    \subfigure[Plot of the average distance $d$ between the two clusters in the KE plot over 30 ad-hoc modular networks. The distance is defined as the euclidean distance between the centroids of the lower half of the cluster.]{
	    \label{fig:distance_ba}
	    \includegraphics[width=0.46\linewidth]{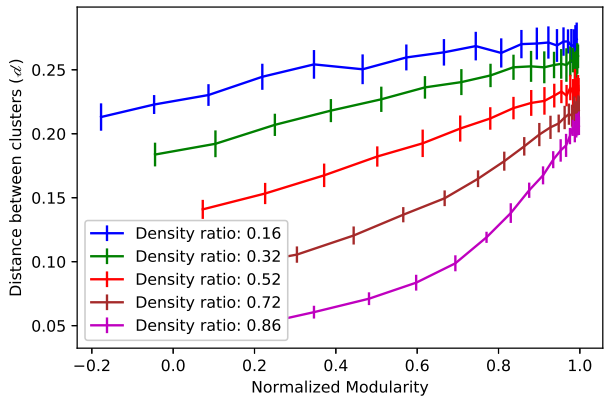}}\qquad
	    \subfigure[Plot of the average ratio of cluster length $\text{\calligra l}_1 / \text{\calligra l}_2$ in the KE plot over 30 ad-hoc modular networks. Cluster length is defined as the length of the diagonal of the smallest rectangle that encapsulates all points of the cluster.]{
	    \label{fig:length_ba}
	    \includegraphics[width=0.46\linewidth]{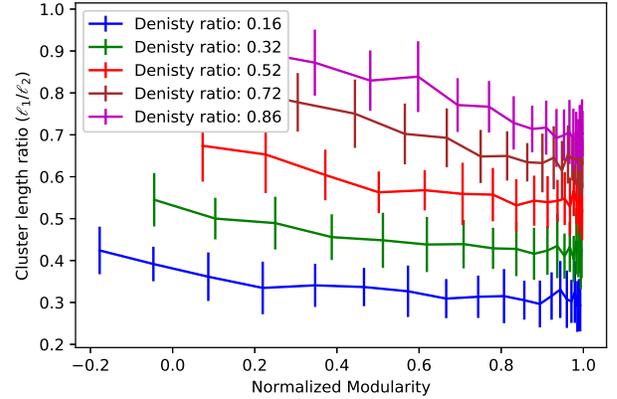}}
	    \caption{Plots regarding the KE plot for ad-hoc modular graphs with the BA model. For \subref{fig:angle_ba}, \subref{fig:distance_ba}, and \subref{fig:length_ba}, $n_1=n_2=250$ remains consistent. Error bars are one standard deviation.}
	    \label{fig:BA_study}
	\end{figure*}
	
	\begin{figure*}
	    \centering
	    \subfigure[]{
	    \label{fig:example_er}
	    \includegraphics[width=0.46\linewidth]{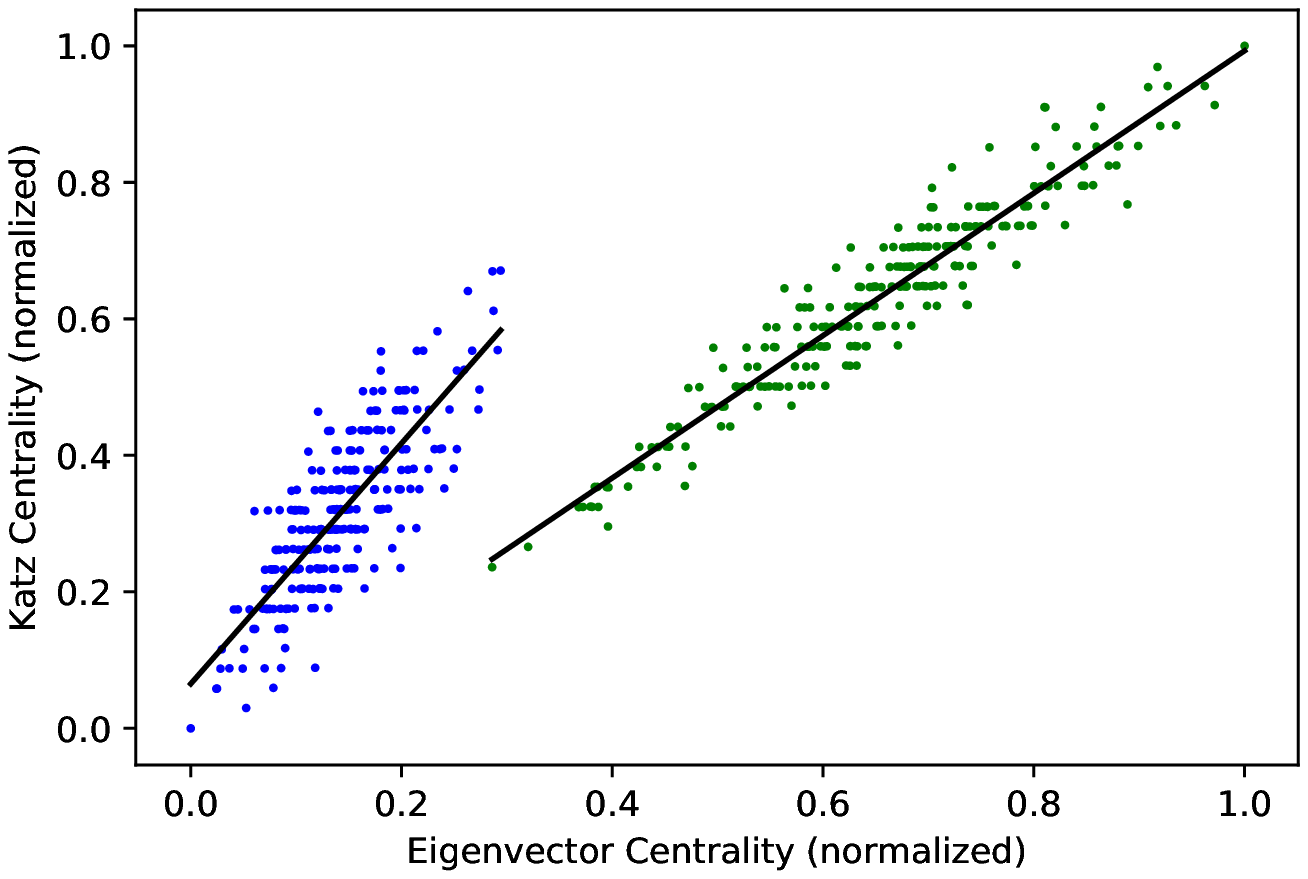}}\qquad
	    \subfigure[]{
	    \label{fig:angle_er}
	    \includegraphics[width=0.46\linewidth]{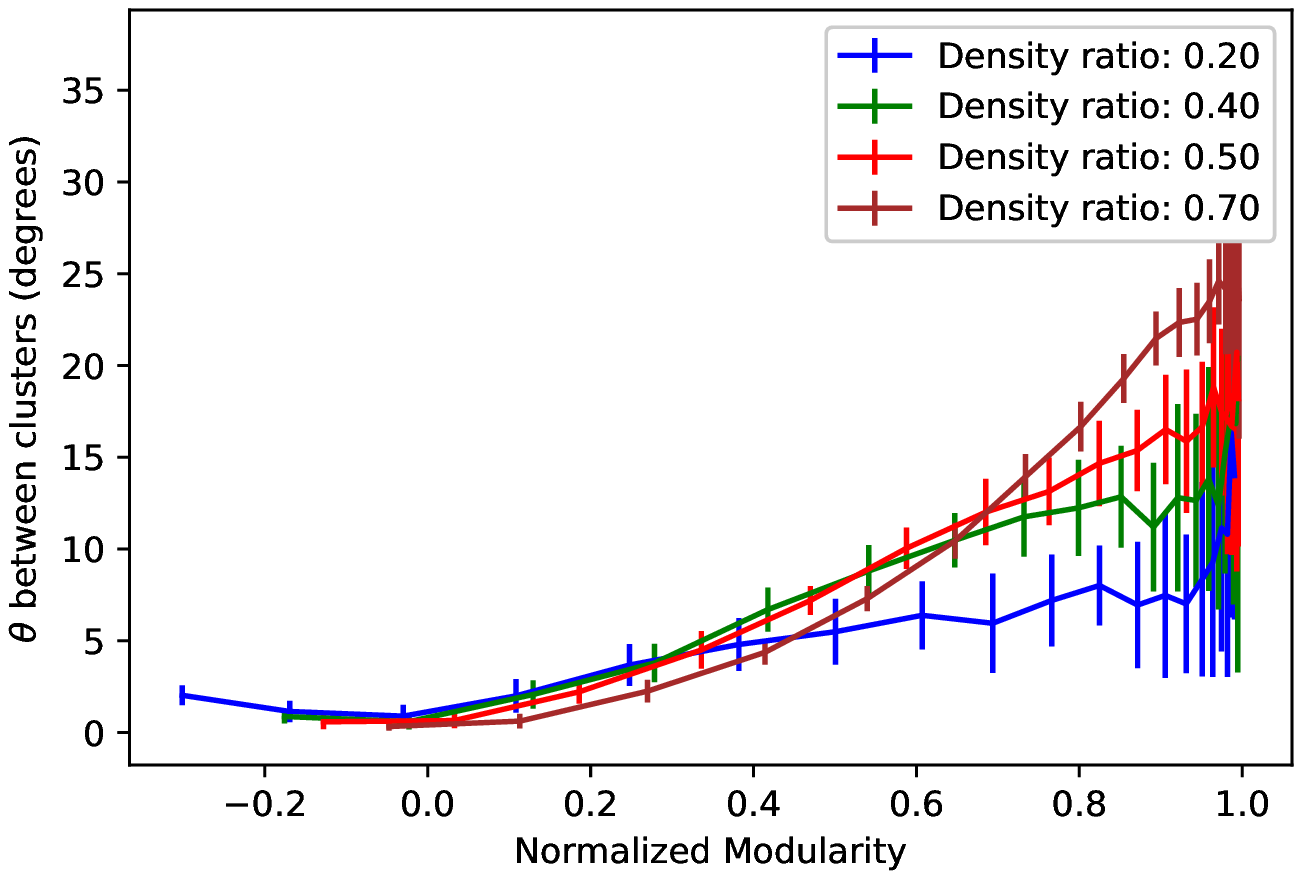}}\\
	    \subfigure[]{
	    \label{fig:distance_er}
	    \includegraphics[width=0.46\linewidth]{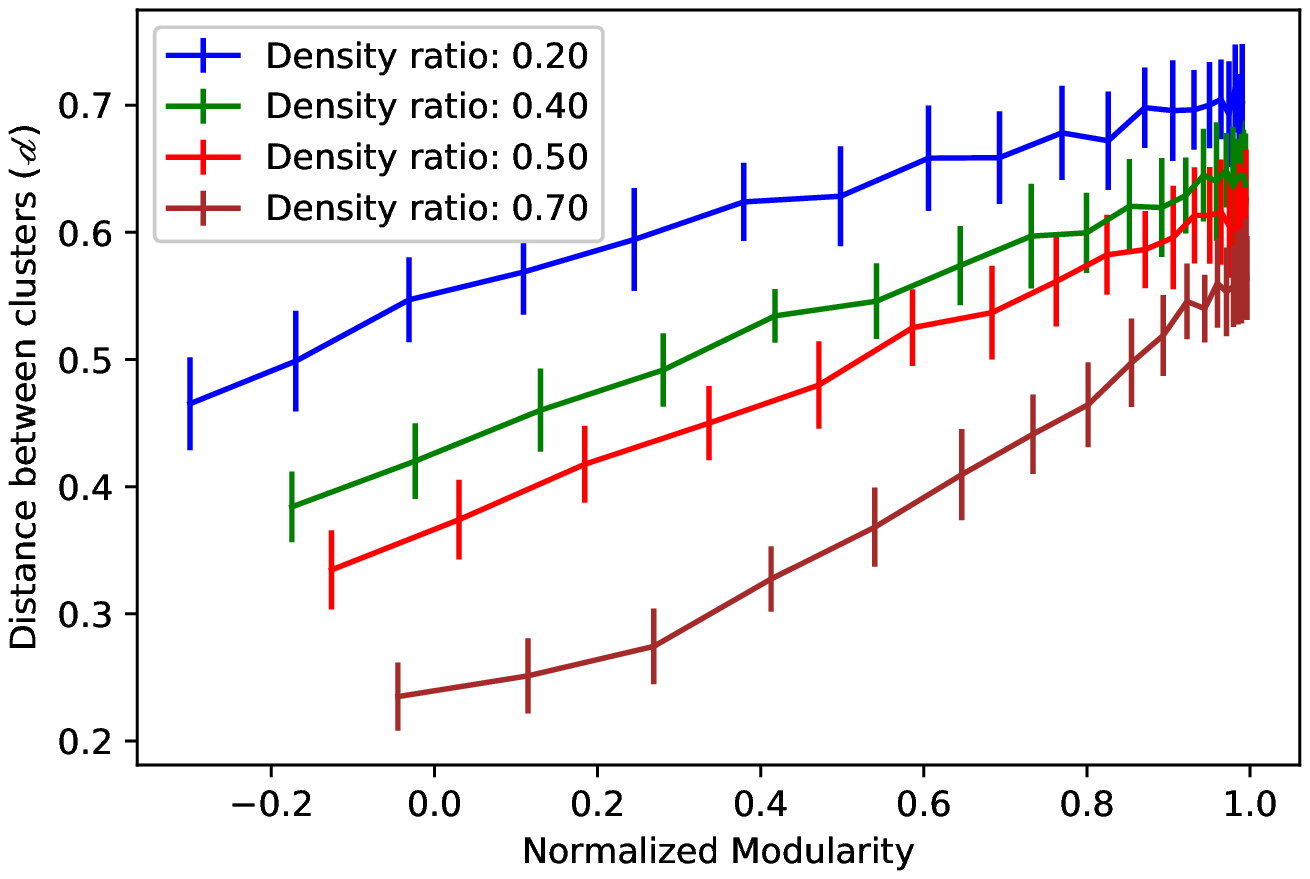}}\qquad
	    \subfigure[]{
	    \label{fig:length_er}
	    \includegraphics[width=0.46\linewidth]{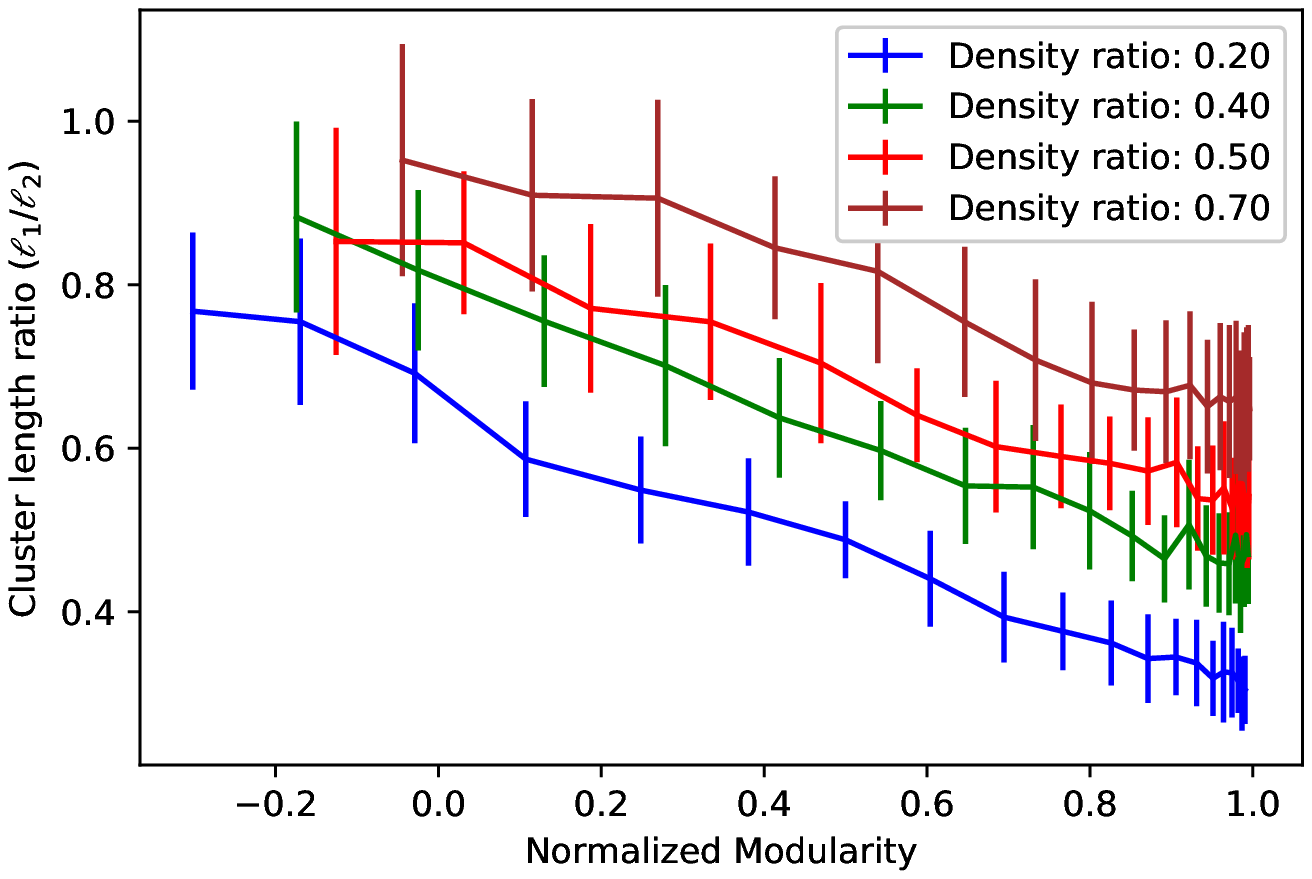}}
	    \caption{The same information given in Figure \ref{fig:BA_study} is given here for ad-hoc modular networks constructed with the ER model.}
	    %5Plots regarding the KE plot for ad-hoc modular graphs with the ER model. \subref{fig:example_er} illustrates an example of the KE plot of one such modular graph. $n_1=n_2=250$ and $\mu=800$. \subref{fig:angle_er} is the plot of the average angle $\theta$ between the two clusters in the KE plot over 20 ad-hoc modular networks. $\theta$ is calculated by finding the difference in arctangent of the linear slope of each cluster. \subref{fig:distance_er} is the plot of the average distance between the two clusters in the KE plot over 20 ad-hoc modular networks. The distance is defined as the euclidean distance between the centroids of the lower half of the cluster. This definition causes the distance to be measured closer to the base of each cluster. \subref{fig:length_er} is the plot of the average ratio of cluster length in the KE plot over 20 ad-hoc modular networks. Cluster length is defined as the length of the diagonal of the smallest rectangle that encapsulates all points of the cluster. For \subref{fig:angle_ba}, \subref{fig:distance_ba}, and \subref{fig:length_ba}, $n_1=n_2=250$ remains consistent.}
	    \label{fig:ER_study}
	\end{figure*}

	The ad-hoc modular networks are formed as follows. Two random graphs are generated with $n_1$ and $n_2$ nodes and $m_1$ and $m_2$ edges, respectively. In this work we use the Erd\H{o}s-R\'enyi (ER) model \cite{ER} and the Barab\'asi-Albert (BA) model \cite{BA}. 
	Each of these models are tunable by a single parameter (ER - the probability $p$ of edge existance; BA - the number of edges $q$ added by each node) and we use these to control the density of the networks.
	%Adjusting the $p$ parameter (the probability of edge existance) for the ER model and the $m$ parameter (the number of edges added by each new node) for the BA model controls the density of each random network.
	
	The two networks are then joined by adding $\mu$ edges between (uniformly) randomly selected nodes from each, thereby creating one network of two communities. Increasing $\mu$ decreases the modularity of the network and vice versa.
	
	Figure \ref{fig:example_ba} shows the plot of the normalized Katz vs eigenvector centrality of ad-hoc modular networks. When calculating Katz centrality, $\alpha$ and $\beta$ are set to 0.001 and 1, respectively. In order to allow for comparison across different networks, Katz and eigenvector centralities are normalized to be between $[0,1]$. Because of this normalization, as long as $\alpha < \lambda_1$, the $\alpha$ and $\beta$ values make no difference in the resulting KE plot.
	
	The green community is created to be more dense than the blue, and there are clear differences in the KE plot. A linear regression of each community shows differing slope, and there is a shift in the lowest eigenvector centrality and Katz centrality of each community.
	
	Performing parameter sweeps of $p$ (for ER) or $q$ (for BA) and  of $\mu$ allows for the characterization of the KE plot across varying modularities and differences in density between the communities.
	We calculate the density of a community by
	\begin{equation}
	    \rho_k = \frac{2m_k}{n_k(n_k-1)}~,~ k=1,2,\dots~ ,
	\end{equation}
	the fraction of the possible edges that exist within the $k$th community.
	When the ad-hoc networks are created, the density of each subnetwork is calculated before they are joined. Figures \ref{fig:angle_ba} - \subref{fig:length_ba} refer to the density ratio. This is the ratio for the density of the sparser (blue) community to that of the denser (green) community. As this ratio gets closer to 1, it is understood that the densities of the communities become more similar.
	
	Figure \ref{fig:angle_ba} shows that since the angle $\theta$ between the clusters increases with modularity, higher assortative mixing in the network is associated with more separability in the KE plot. As the density ratio of the communities approach equivalency, $\theta$ becomes more sensitive to modularity. A near-linear relationship at lower density ratios turns more quadratic at higher density ratios. At low density ratios and high modularity, the sparse cluster on the KE plot loses its clear linear trend, causing the linear regression to perform erratically and subsequently causes the increased variance in $\theta$ seen in Figures \ref{fig:angle_ba}-\subref{fig:length_ba} at high modularity and low density ratio.
	
	In addition to $\theta$ changing with modularity and density ratio, the euclidean distance between the bases of the two clusters, $d$, changes as well. Figure \ref{fig:distance_ba} shows $d$ increasing with stronger assortative mixing and decreasing with higher density ratio. Again, we see the higher sensitivity to modularity when the density ratio approaches 1.
	
	The relative size of the dense and sparse clusters is also dependent on modularity and density. Figure \ref{fig:length_ba} shows the cluster length ratio between the communities: $\text{\calligra l}_1 / \text{\calligra l}_2$. As modularity increases, $\text{\calligra l}_1 / \text{\calligra l}_2$ decreases. The closer the communities' densities are to each other, the higher modularity required to achieve a similar difference between $\text{\calligra l}_1$ and  $\text{\calligra l}_2$.
	
	The fact that differences in community density, connectivity, and modularity result in discernible differences in the KE plot suggests that this plot can be useful for community detection in networks. Similar outcomes for ad-hoc networks constructed with the ER model (see Figure \ref{fig:ER_study}) indicate that networks with exponential or scale-free degree distributions are both apt to demonstrate this phenomenon.
	
\section{Performance on Real World Networks}
	\label{s:realWorld}
	To further test this proposed community detection method we look at real-world networks. Networks with predetermined ``ground-truth" communities are especially of interest since they can provide a measurable benchmark. However, it should be noted that the existence of ``ground-truth" communities may not translate to a strong assortatively mixed network. For example, in a network of United States congressional seats based on vote-sharing, we can define ``ground-truth" communities by gender or age. But the modularity due to these communities will not out-perform that caused by communities of political party \cite{congress}.
	
	One network with predetermined ``ground-truth" communities is the DBLP network \cite{DBLP} obtained from the Stanford Network Analysis Project \cite{SNAP}. The DBLP is a co-authorship network of computer science publications. ``Publication venue, e.g, journal or conference, defines an individual ground-truth community; authors who published to a certain journal or conference form a community."
	
\begin{figure*}
	    \centering
	    \subfigure[DBLP network - ground truth]{
	    \label{fig:dblp_pre}
	    \includegraphics[width=0.4\linewidth]{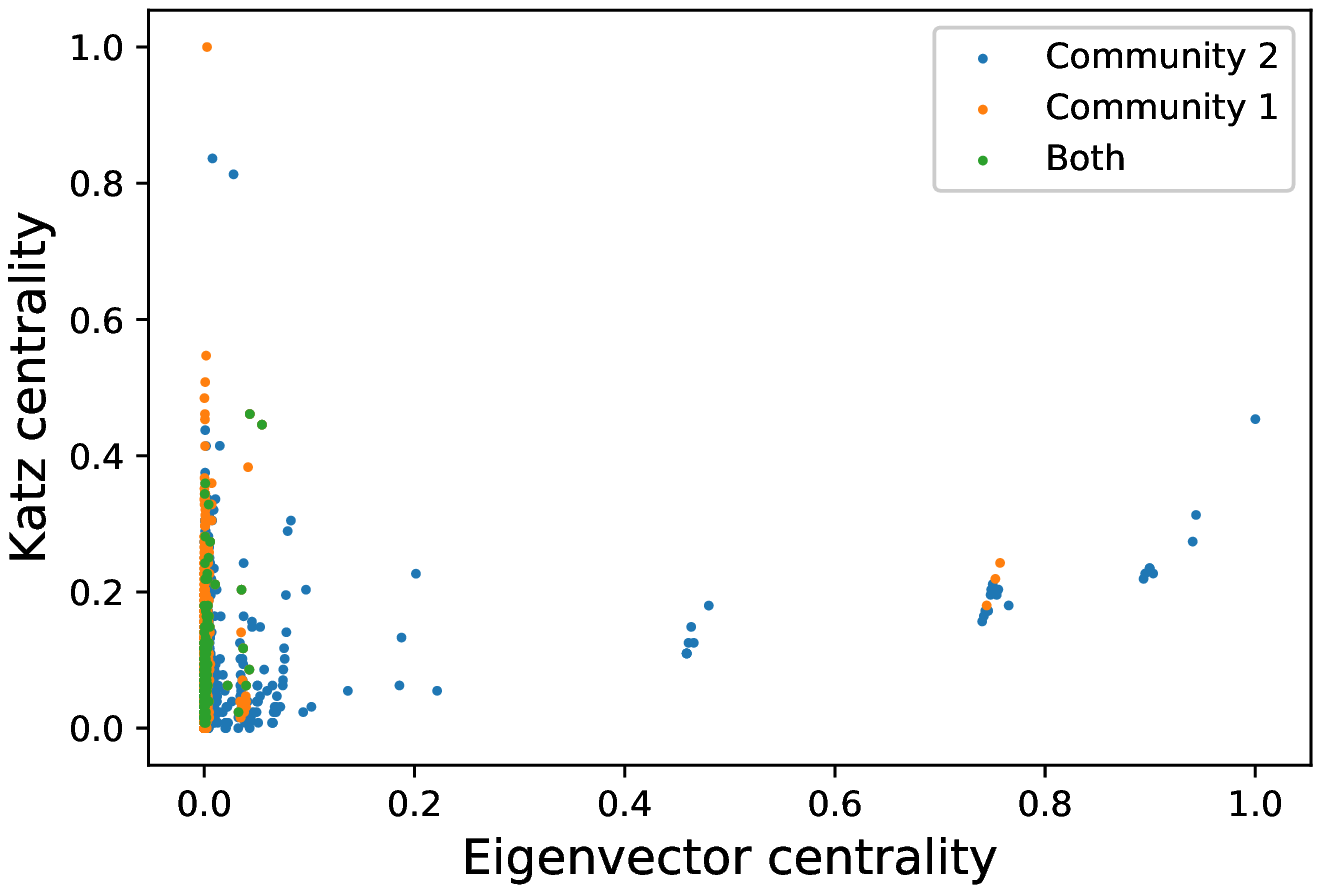}}\qquad
    	\subfigure[DBLP network - proposed method]{
    	\label{fig:dblp_cluster}
    	\includegraphics[width=0.4\linewidth]{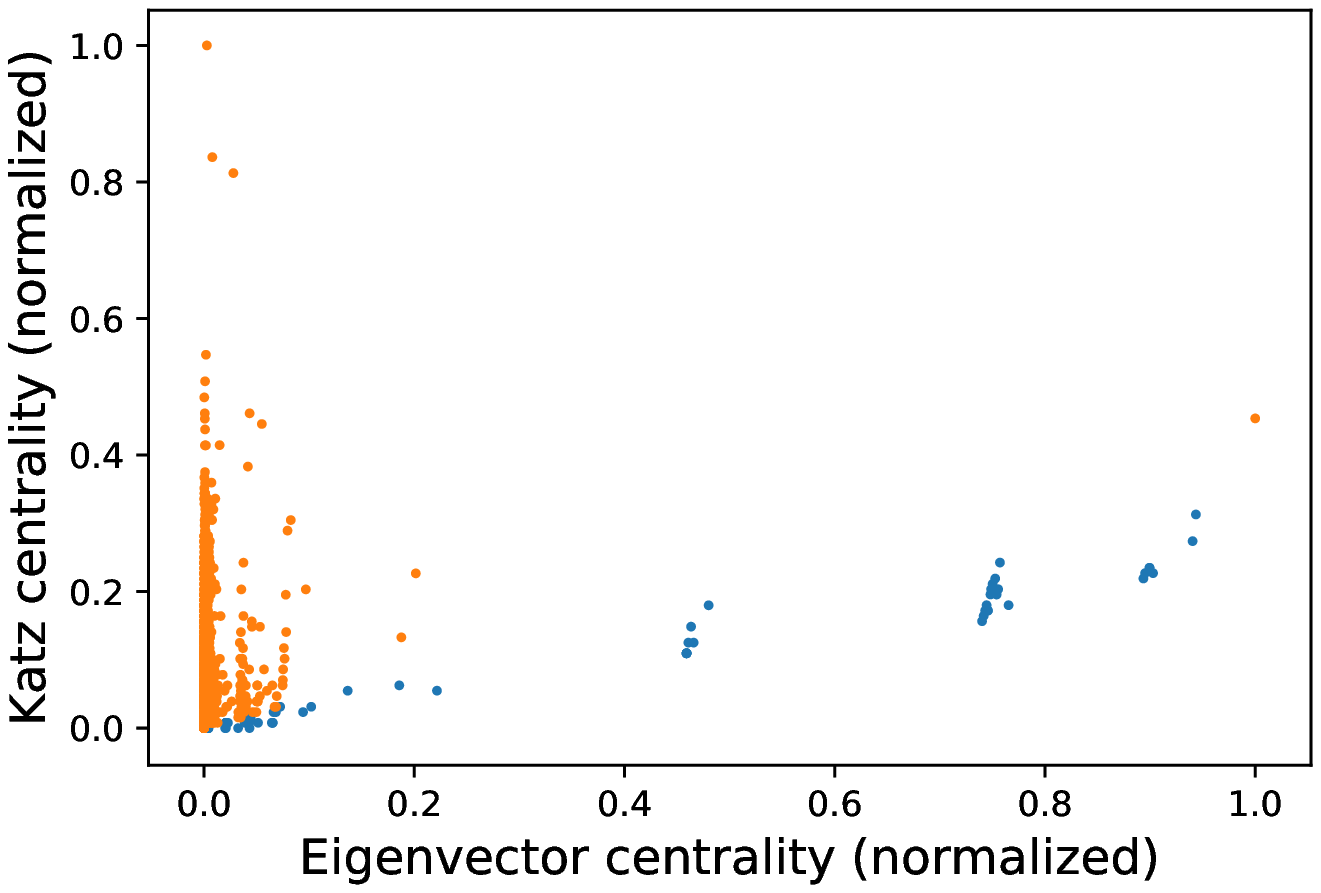}}\\ \vspace{0.1cm}
    	\subfigure[YouTube friendship network - ground truth]{
    	\label{fig:youtube}
    	\includegraphics[width=0.4\linewidth]{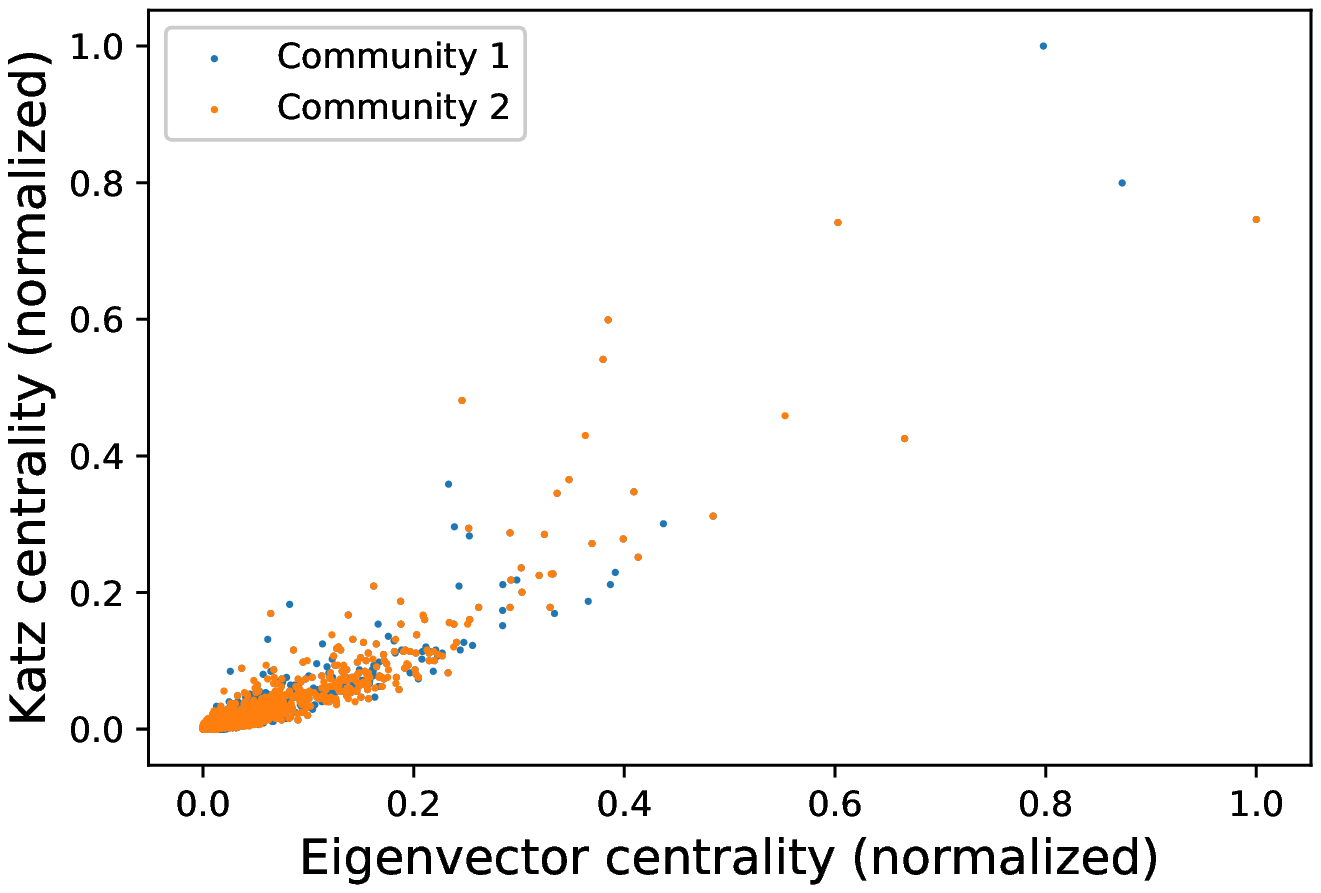}}\qquad
    	\subfigure[Amazon product network - proposed method]{
    	\label{fig:amazon_product}
    	\includegraphics[width=0.4\linewidth]{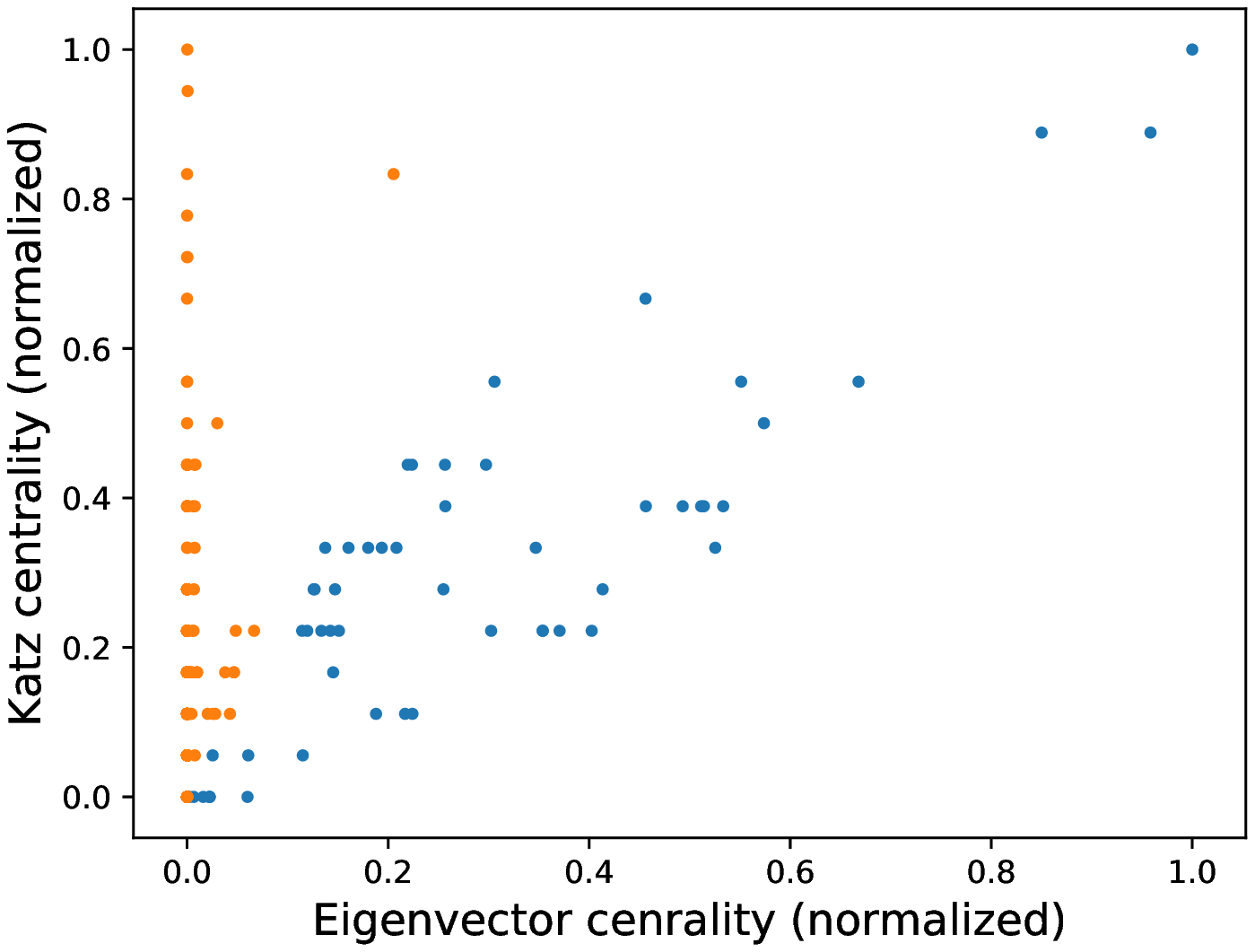}}\\ \vspace{0.1cm}
    	\subfigure[Amazon health network - proposed method]{
    	\label{fig:amazon_health}
    	\includegraphics[width=0.4\linewidth]{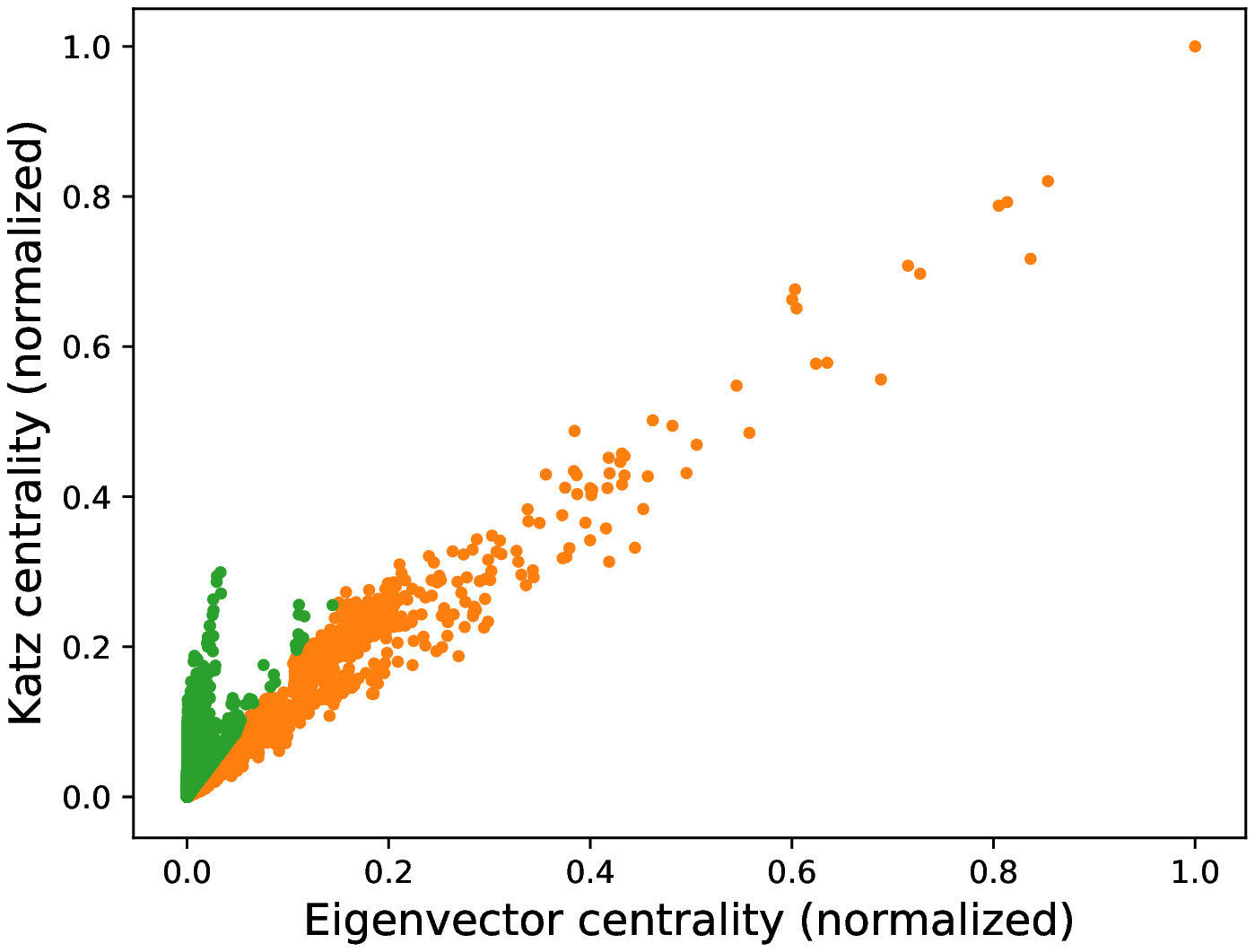}}\qquad
    	\subfigure[Amazon beauty network - proposed method]{
    	\label{fig:amazon_beauty}
    	\includegraphics[width=0.4\linewidth]{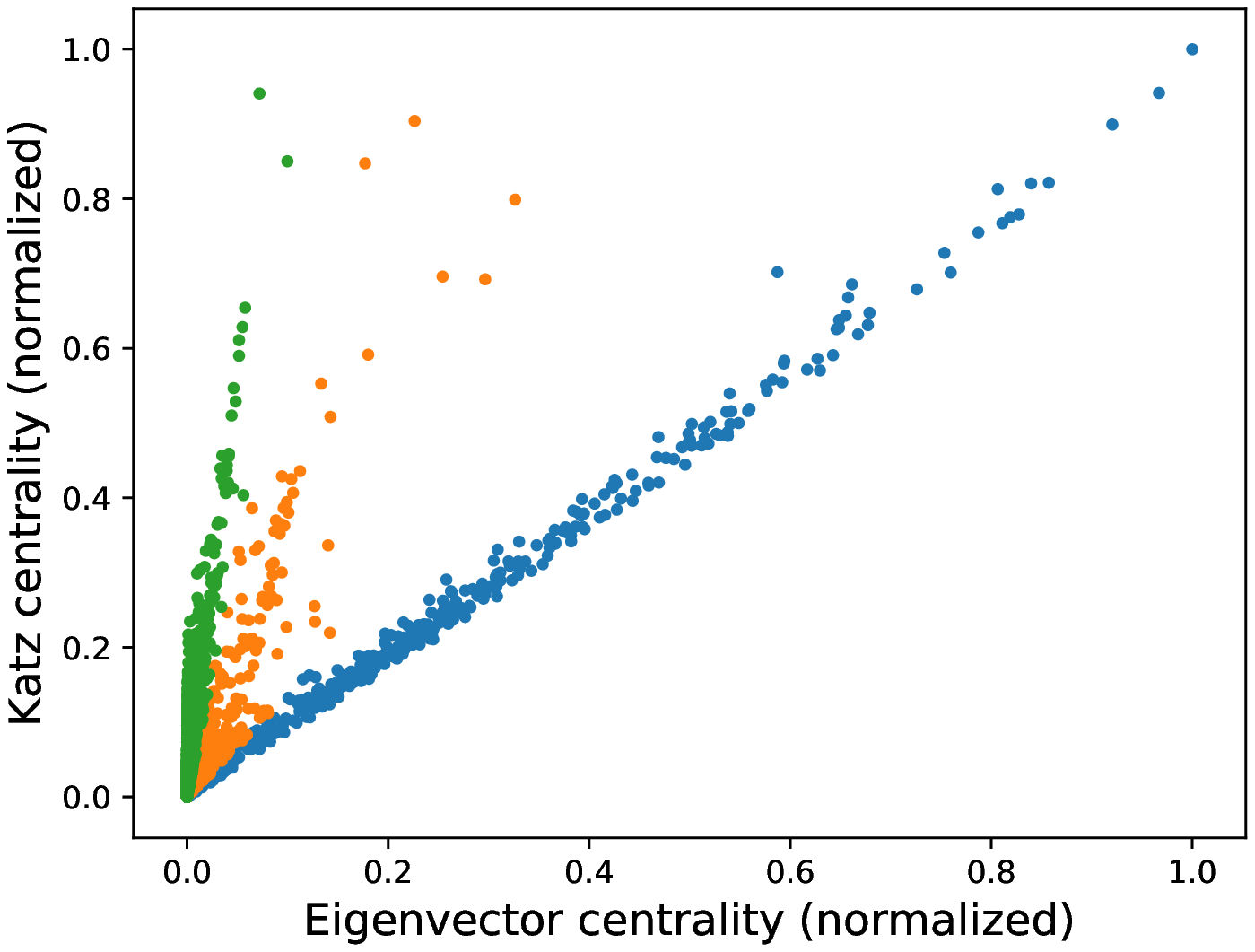}}
    	\caption{\subref{fig:dblp_pre} KE plot of the sampled DBLP network. For Katz centrality, $\alpha = 1e-4$ and $\beta = 1$. \subref{fig:dblp_cluster} KE plot of the sampled DBLP clustering showing the clusters found by the clustering algorithm. \subref{fig:youtube} KE plot of two largest communities in the YouTube friendship network. This sampled graph has 3,269 nodes, and 25\% of nodes are members of both communities. \subref{fig:amazon_product} KE plot of the two largest communities in the Amazon product network. The sampled graph has 328 nodes, with all nodes belonging to both ``ground-truth" communities. The blue and orange clusters are the detected communities. \subref{fig:amazon_health} The KE plot of the Amazon health product review network. The orange and green clusters are the detected communities. \subref{fig:amazon_beauty} The KE plot of the Amazon beauty product review network. The blue, orange, and green clusters are the detected communities.}
    	\label{fig:all_examples}
	\end{figure*}

	There are 317,080 nodes in over 5,000 ``ground truth" communities in this network. We sample the nodes in the two largest communities and take the induced subgraph, leaving our network with 13,326 nodes. Since communities are defined from publication venue, it is possible for nodes to belong to multiple communities. The KE plot is shown in Figure \ref{fig:dblp_pre}. The orange and blue dots represent the nodes in communities one and two, respectively, and the green dots are nodes that are in both.
	
	We see two distinct vectors in the KE plot: one dominated by primarily blue nodes, and the other dominated by orange and green. The wide angle between them suggests high modularity. In fact, between the two distinct communities, $Q=0.438$, and $Q/Q_{max} = 0.827$.
	
	Assume that these labels were not provided before hand. Using our clustering method, we can define two communities of nodes in the KE plot. The results of this clustering is shown in Figure \ref{fig:dblp_cluster}. Assuming these two node groupings, $Q=0.019$ and $Q/Q_{max}=0.567$. The modularity is 70\% of that seen with the ``ground-truth" community labels. $Q$ is so low here due to the imbalance in cluster size. The orange cluster consists of only 462 nodes, whereas the blue cluster contains the remaining 12,864 nodes. Thus, only a small fraction of the edges can be contained in the orange cluster.
	
	As suggested by the trends in Section \ref{s:randomGraphs}, lack of modular communities results in little-to-no distinguishable clusters in the KE plot. This is illustrated in a YouTube friendship network, where ``ground-truth" communities are defined by user groups \cite{DBLP,youtube}. The KE plot of this network is shown in Figure \ref{fig:youtube}. The significant overlap of the two communities on the plot indicates low modularity: $Q=0.049$ and $Q/Q_{max}=0.109$. In this case, the ``ground-truth" communities do not reflect the properties of communities from a network perspective.

	Similarly, two largest ``ground-truth" communities in the Amazon product network \cite{youtube,DBLP} contain the same nodes. As a result, $Q=0$ and the KE plot of the two communities are completely overlapped. These two defined communities are not reflected strongly in the network structure. But, as depicted in Figure \ref{fig:amazon_product}, there are two clusters of nodes identified. Using these communities, where $Q=0.324$ and $Q/Q_{max}=0.796$, provides more insight into network-relevant groupings compared to the ``ground-truth" communities.
	
	Many, if not most, networks do not contain ``ground truth" communities, hence the need for community detection algorithms. We now show the applicability of the KE community detection method to other real-world networks without ``ground-truth" communities, and judge its effectiveness solely by the normalized modularity achieved from the detected communities.
	
	We sample the 5-core dataset of Amazon health product reviews \cite{amazon} from January to July of 2014. A review network is constructed where nodes (user accounts) are connected with weighted edges representing the number of common products reviewed. The resulting network has 25,026 nodes. Figure \ref{fig:amazon_health} shows the KE plot of this network after applying the clustering algorithm to identify communities. With the assigned communities modularity is quite high: $Q=0.372$ and $Q/Q_{max}=0.777$. There are no ``ground-truth" communities given for this network, but the high normalized modularity indicates good detection of communities.
	
	The KE community detection method has shown the ability to detect more than two communities. Figure \ref{fig:amazon_beauty} shows the KE plot of another Amazon review network of beauty products (created the same way as the health product network). Here, there are three clear clusters, indicating three communities in the 13,043 node network. With these defined communities, $Q=0.202$ and $Q/Q_{max}=0.645$. A summary of these findings is shown in Table \ref{tab:performance}.
	
	\begin{table}[t]
		\centering
		\begin{tabular}{lcc}
			\toprule
			\textbf{Network} & \textbf{G.T.} $Q/Q_{max}$ & \textbf{Detected $Q/Q_{max}$}\\
			\midrule
			DBLP & 0.438/0.530 & 0.019/0.034\\
			YouTube & 0.049/0.450 & N/A\\
			AMZN Product & 0.000 & 0.324/0.407\\
			AMZN Health & N/A & 0.372/0.479\\
			AMZN Beauty & N/A & 0.202/0.313\\
			\bottomrule
		\end{tabular}
		\caption{Summary of the performance of KE community detection on real-world networks, compared to ground-truth (G.T.) when possible.}
		\label{tab:performance}
	\end{table}
	
    \begin{table*}
        \centering
        \begin{tabular}{lrrrr}
            \toprule
            & \multicolumn{2}{c}{\textbf{Runtime (sec)}} & \multicolumn{2}{c}{\textbf{Modularity ($Q/Q_{max}$)}}\\
            \midrule
            \textbf{Network $(n,m)$} & \textbf{Louvain} & \textbf{KE} & \textbf{Louvain} & \textbf{KE}\\
            \midrule
            AMZN Prod. (328, 679)   & 0.371 sec & 0.032 sec & 0.801/0.908 & 0.359/0.467 \\
            AdHoc BA (1000, 16975)  & 11.8 sec  & 0.329 sec & 0.485/0.491 & 0.480/0.498 \\
            AdHoc BA (2000, 11035)  & 122 sec   & 0.228 sec & 0.291/0.930 & 0.228/0.471 \\
            AdHoc BA (10000,95079)  & 7,452 sec & 1.97 sec  & 0.203/0.931 & 0.123/0.492 \\
            \bottomrule
        \end{tabular}
        \caption{Comparison of wall time and resulting modularity between Louvain community detection and the KE plot method of extracting communities.}
        \label{tab:quality}
    \end{table*}
	
\section{On the quality of the detected communities}
\label{s:qual}
    We evaluate the quality of the KE community detection method from two perspectives: the modularity and speed compared to the Louvain method.
    
    \begin{figure*}
	    \centering
	    \subfigure[]{
	    \label{fig:louvain1}
	    \includegraphics[width=0.48\linewidth]{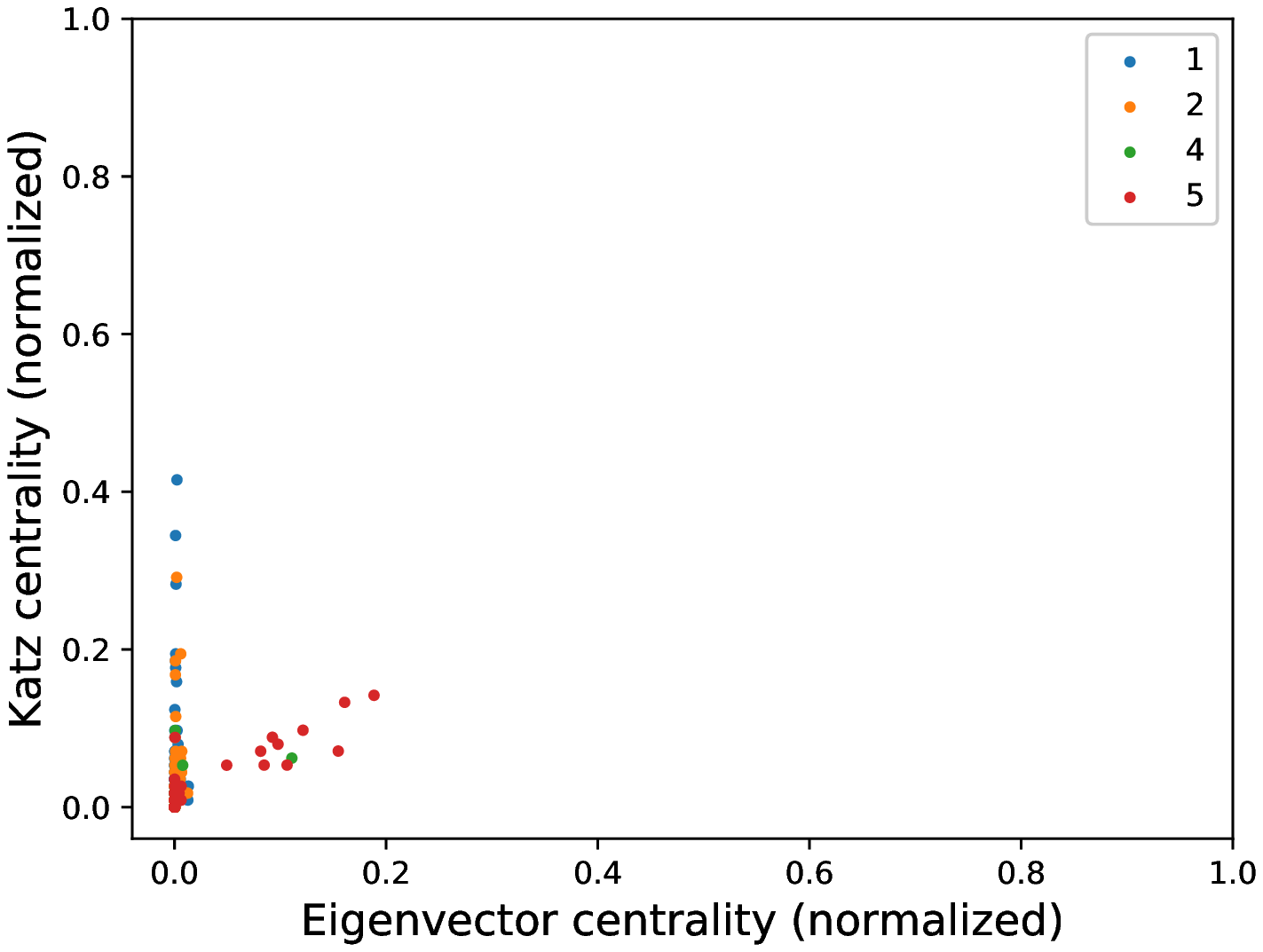}}
	    \subfigure[]{
	    \label{fig:louvain2}
	    \includegraphics[width=0.48\linewidth]{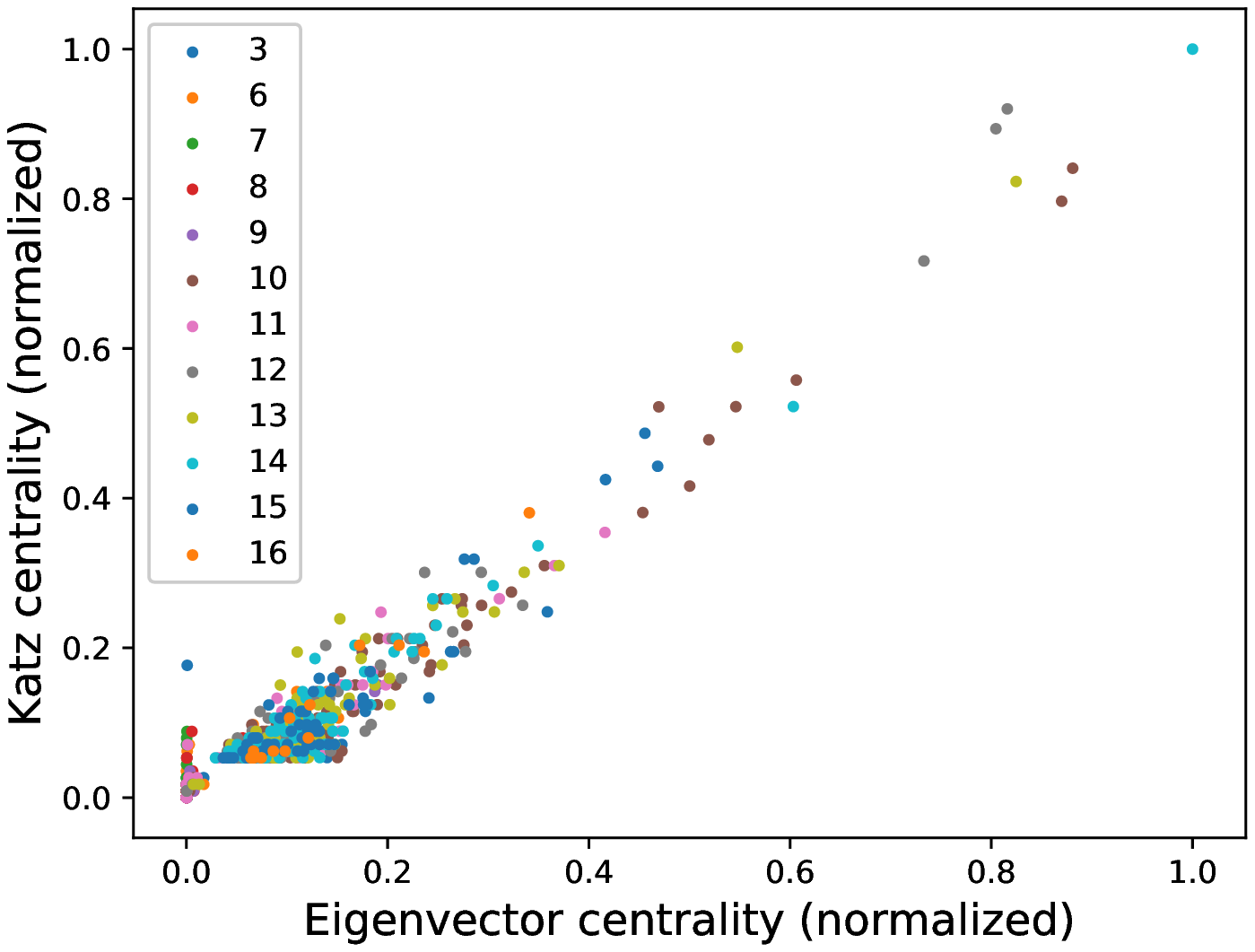}}
	    \caption{KE plots of a 1,000-node/5,032-edge ad-hoc modular network group by communities detected from Louvain showing \subref{fig:louvain1} the first four communities and \subref{fig:louvain2} the remaining seven communities. A few outliers notwithstanding, we see that the two groupings follow the common pattern observed in the KE plots of modular networks.}
	    \label{fig:louvain_ex}
	\end{figure*}
    
    Figure \ref{fig:louvain_ex} shows two KE plots of an ad-hoc modular network. Louvain community detection was performed, finding sixteen communities. However, as shown in Figures \ref{fig:louvain1} and \ref{fig:louvain2}, these sixteen communities can be combined into two communities (1,2,4,5 and the remaining) that mostly display the previously exhibited trends in the KE plot. With the eleven communities detected from Louvain, $Q=0.228$ and $Q/Q_{max}=0.299$. By taking the two larger communities suggested by the KE plots in Figure \ref{fig:louvain_ex}, and reassigning the outlier nodes, we have $Q=0.309$ and $Q/Q_{max} = 0.940$; slightly higher modularity, but much higher normalized modularity than the communities resulting from Louvain. We take as validation the fact that the communities found using our KE approach both align with the communities found by Louvain and also that the proposed method provides increased modularity scores.
    
    %\textcolor{red}{That Louvain community detection results in communities supportive of the relationship between the KE plot and modularity-based community detection is \textellipsis and something about how a better unsupervised clustering algorithm could be better.}
	
	The superior speed of the KE method is another capability of the proposed method. Table \ref{tab:quality} gives a side-by-side comparison of the wall times and modularities for Louvain  community detection and the KE method on the real-world Amazon product network and three ad-hoc modular networks. Louvain calculation time grows at a much faster rate with respect to node/edge number. For large networks with millions of nodes, a highly distributed compute infrastructure is needed to perform Louvain community detection in reasonable time \cite{louvainDist}. Conversely, the KE method will continue to be able to perform on the order of seconds for large scale networks. Regarding modularity, both methods result in comparable modularities or normalized modularities, though some of the examples show the Louvain method performing better. The simplistic clustering algorithm may limit the performance of the KE method in this study, as the distinction between the clusters tends to become less clear near the origin.
	
\section{Conclusions}
	\label{s:conc}
	In this work, we have showed that Katz and eigenvector centrality provide strong indicators of modularity in networks and demonstrate how this fact can be leveraged for community detection. For undirected networks with sufficient modularity, the KE plot displays clusters aligned along distinct vectors. Clustering algorithms, such as the Radon transform approach used here, can then identify these clusters to assign community membership to each node.
	
	This method was shown to detect communities with comparable modularity to that of pre-labeled ``ground truth" communities in real world networks. In other real world networks without known node labels, the communities detected by the KE method were strongly assortatively mixed, suggesting well-performing community detection. In comparison to the existing Louvain community detection algorithm, we saw the runtime of the KE method as low as $0.02\%$ of the Louvain runtime. Additionally, we saw that communities detected by the Louvain method support the KE plot representation of communities. And when the KE plot is used to refine the Louvain communities, modularity increases.
	
	Future work will investigate alternative clustering algorithms to better extract communities from the KE plot. One attractive method worth exploring is using the KE method to find an initial grouping of nodes to be improved upon by more traditional modularity maximization methods. Furthermore, the impact of edge weights on this method should be more closely understood, particularly for signed networks as well as directed networks.

\appendix


\begin{thebibliography}{26}%
\makeatletter
\providecommand \@ifxundefined [1]{%
 \@ifx{#1\undefined}
}%
\providecommand \@ifnum [1]{%
 \ifnum #1\expandafter \@firstoftwo
 \else \expandafter \@secondoftwo
 \fi
}%
\providecommand \@ifx [1]{%
 \ifx #1\expandafter \@firstoftwo
 \else \expandafter \@secondoftwo
 \fi
}%
\providecommand \natexlab [1]{#1}%
\providecommand \enquote  [1]{``#1''}%
\providecommand \bibnamefont  [1]{#1}%
\providecommand \bibfnamefont [1]{#1}%
\providecommand \citenamefont [1]{#1}%
\providecommand \href@noop [0]{\@secondoftwo}%
\providecommand \href [0]{\begingroup \@sanitize@url \@href}%
\providecommand \@href[1]{\@@startlink{#1}\@@href}%
\providecommand \@@href[1]{\endgroup#1\@@endlink}%
\providecommand \@sanitize@url [0]{\catcode `\\12\catcode `\$12\catcode
  `\&12\catcode `\#12\catcode `\^12\catcode `\_12\catcode `\%12\relax}%
\providecommand \@@startlink[1]{}%
\providecommand \@@endlink[0]{}%
\providecommand \url  [0]{\begingroup\@sanitize@url \@url }%
\providecommand \@url [1]{\endgroup\@href {#1}{\urlprefix }}%
\providecommand \urlprefix  [0]{URL }%
\providecommand \Eprint [0]{\href }%
\providecommand \doibase [0]{https://doi.org/}%
\providecommand \selectlanguage [0]{\@gobble}%
\providecommand \bibinfo  [0]{\@secondoftwo}%
\providecommand \bibfield  [0]{\@secondoftwo}%
\providecommand \translation [1]{[#1]}%
\providecommand \BibitemOpen [0]{}%
\providecommand \bibitemStop [0]{}%
\providecommand \bibitemNoStop [0]{.\EOS\space}%
\providecommand \EOS [0]{\spacefactor3000\relax}%
\providecommand \BibitemShut  [1]{\csname bibitem#1\endcsname}%
\let\auto@bib@innerbib\@empty
%</preamble>
\bibitem [{\citenamefont {Lu}\ \emph {et~al.}(2015)\citenamefont {Lu},
  \citenamefont {Sun}, \citenamefont {Wen}, \citenamefont {Cao},\ and\
  \citenamefont {Porta}}]{DTN}%
  \BibitemOpen
  \bibfield  {author} {\bibinfo {author} {\bibfnamefont {Z.}~\bibnamefont
  {Lu}}, \bibinfo {author} {\bibfnamefont {X.}~\bibnamefont {Sun}}, \bibinfo
  {author} {\bibfnamefont {Y.}~\bibnamefont {Wen}}, \bibinfo {author}
  {\bibfnamefont {G.}~\bibnamefont {Cao}}, and\ \bibinfo {author}
  {\bibfnamefont {T.~L.}\ \bibnamefont {Porta}},\ }\bibfield  {title} {\bibinfo
  {title} {Algorithms and applications for community detection in weighted
  networks},\ }\href@noop {} {\bibfield  {journal} {\bibinfo  {journal} {IEEE
  Transactions on Parallel and Distributed Systems}\ }\textbf {\bibinfo
  {volume} {26}} (\bibinfo {year} {2015})}\BibitemShut {NoStop}%
\bibitem [{\citenamefont {Nguyen}\ \emph {et~al.}(2010)\citenamefont {Nguyen},
  \citenamefont {Xaun},\ and\ \citenamefont {Thai}}]{worm1}%
  \BibitemOpen
  \bibfield  {author} {\bibinfo {author} {\bibfnamefont {N.~P.}\ \bibnamefont
  {Nguyen}}, \bibinfo {author} {\bibfnamefont {Y.}~\bibnamefont {Xaun}}, and\
  \bibinfo {author} {\bibfnamefont {M.~T.}\ \bibnamefont {Thai}},\ }\bibfield
  {title} {\bibinfo {title} {A novel method for worm containment on dynamic
  social networks},\ }in\ \href@noop {} {\emph {\bibinfo {booktitle}
  {MILCOM}}}\ (\bibinfo {year} {2010})\BibitemShut {NoStop}%
\bibitem [{\citenamefont {Peixoto}(2019)}]{reconstruct}%
  \BibitemOpen
  \bibfield  {author} {\bibinfo {author} {\bibfnamefont {T.~P.}\ \bibnamefont
  {Peixoto}},\ }\href@noop {} {\bibinfo {title} {Network reconstruction and
  community detection from dynamics}},\ \bibinfo {howpublished}
  {arXiv:1903.10833 [physics.soc-ph]} (\bibinfo {year} {2019})\BibitemShut
  {NoStop}%
\bibitem [{\citenamefont {Girvan}\ and\ \citenamefont {Newman}(2004)}]{Mod}%
  \BibitemOpen
  \bibfield  {author} {\bibinfo {author} {\bibfnamefont {M.}~\bibnamefont
  {Girvan}}and\ \bibinfo {author} {\bibfnamefont {M.~E.~J.}\ \bibnamefont
  {Newman}},\ }\bibfield  {title} {\bibinfo {title} {Finding and evaluating
  community structure in networks},\ }\href@noop {} {\bibfield  {journal}
  {\bibinfo  {journal} {Phys. Rev. E}\ } (\bibinfo {year} {2004})}\BibitemShut
  {NoStop}%
\bibitem [{\citenamefont {Brandes}\ \emph {et~al.}(2008)\citenamefont
  {Brandes}, \citenamefont {Delling}, \citenamefont {Gaertler}, \citenamefont
  {Gorke}, \citenamefont {Hoefer}, \citenamefont {Nikoloski},\ and\
  \citenamefont {Wagner}}]{NP}%
  \BibitemOpen
  \bibfield  {author} {\bibinfo {author} {\bibfnamefont {U.}~\bibnamefont
  {Brandes}}, \bibinfo {author} {\bibfnamefont {D.}~\bibnamefont {Delling}},
  \bibinfo {author} {\bibfnamefont {M.}~\bibnamefont {Gaertler}}, \bibinfo
  {author} {\bibfnamefont {R.}~\bibnamefont {Gorke}}, \bibinfo {author}
  {\bibfnamefont {M.}~\bibnamefont {Hoefer}}, \bibinfo {author} {\bibfnamefont
  {Z.}~\bibnamefont {Nikoloski}}, and\ \bibinfo {author} {\bibfnamefont
  {D.}~\bibnamefont {Wagner}},\ }\bibfield  {title} {\bibinfo {title} {On
  modularity clustering},\ }\href@noop {} {\bibfield  {journal} {\bibinfo
  {journal} {IEEE Transactions on Knowledge and Data Engineering}\ } (\bibinfo
  {year} {2008})}\BibitemShut {NoStop}%
\bibitem [{\citenamefont {Newman}(2006{\natexlab{a}})}]{simpleMM2}%
  \BibitemOpen
  \bibfield  {author} {\bibinfo {author} {\bibfnamefont {M.~E.~J.}\
  \bibnamefont {Newman}},\ }\bibfield  {title} {\bibinfo {title} {Modularity
  and community structure in newtworks},\ }\href@noop {} {\bibfield  {journal}
  {\bibinfo  {journal} {Proc. Natl. Acad. Sci.}\ }\textbf {\bibinfo {volume}
  {103}} (\bibinfo {year} {2006}{\natexlab{a}})}\BibitemShut {NoStop}%
\bibitem [{\citenamefont {Blondel}\ \emph {et~al.}(2008)\citenamefont
  {Blondel}, \citenamefont {Guillaume}, \citenamefont {Lambiotte},\ and\
  \citenamefont {Lefebvre}}]{louvain}%
  \BibitemOpen
  \bibfield  {author} {\bibinfo {author} {\bibfnamefont {V.~D.}\ \bibnamefont
  {Blondel}}, \bibinfo {author} {\bibfnamefont {J.-L.}\ \bibnamefont
  {Guillaume}}, \bibinfo {author} {\bibfnamefont {R.}~\bibnamefont
  {Lambiotte}}, and\ \bibinfo {author} {\bibfnamefont {E.}~\bibnamefont
  {Lefebvre}},\ }\bibfield  {title} {\bibinfo {title} {Fast unfolding of
  communities in large networks},\ }\href@noop {} {\bibfield  {journal}
  {\bibinfo  {journal} {Journal of Statistical Mechanics Theory and
  Experiment}\ } (\bibinfo {year} {2008})}\BibitemShut {NoStop}%
\bibitem [{\citenamefont {Newman}(2006{\natexlab{b}})}]{spectral}%
  \BibitemOpen
  \bibfield  {author} {\bibinfo {author} {\bibfnamefont {M.~E.~J.}\
  \bibnamefont {Newman}},\ }\bibfield  {title} {\bibinfo {title} {Finding
  community structure in networks using the eigenvectors of matrices},\
  }\href@noop {} {\bibfield  {journal} {\bibinfo  {journal} {Phys. Rev. E}\
  }\textbf {\bibinfo {volume} {74}} (\bibinfo {year}
  {2006}{\natexlab{b}})}\BibitemShut {NoStop}%
\bibitem [{\citenamefont {Traag}\ \emph {et~al.}(2019)\citenamefont {Traag},
  \citenamefont {Waltman},\ and\ \citenamefont {van Eck}}]{Leider}%
  \BibitemOpen
  \bibfield  {author} {\bibinfo {author} {\bibfnamefont {V.~A.}\ \bibnamefont
  {Traag}}, \bibinfo {author} {\bibfnamefont {L.}~\bibnamefont {Waltman}}, and\
  \bibinfo {author} {\bibfnamefont {N.~J.}\ \bibnamefont {van Eck}},\
  }\bibfield  {title} {\bibinfo {title} {From louvain to leiden: guaranteeing
  well-connected communities},\ }\href@noop {} {\bibfield  {journal} {\bibinfo
  {journal} {Scientific reports}\ }\textbf {\bibinfo {volume} {9}} (\bibinfo
  {year} {2019})}\BibitemShut {NoStop}%
\bibitem [{\citenamefont {Ghosh}\ and\ \citenamefont {Lerman}(2011)}]{alpha}%
  \BibitemOpen
  \bibfield  {author} {\bibinfo {author} {\bibfnamefont {R.}~\bibnamefont
  {Ghosh}} and\ \bibinfo {author} {\bibfnamefont {K.}~\bibnamefont {Lerman}},\
  }\bibfield  {title} {\bibinfo {title} {A parameterized centrality metric for
  network analysis},\ }\href@noop {} {\bibfield  {journal} {\bibinfo  {journal}
  {Phys. Rev. E}\ }\textbf {\bibinfo {volume} {83}} (\bibinfo {year}
  {2011})}\BibitemShut {NoStop}%
 \bibitem [{\citenamefont {Ditsworth}(2019)}]{github}%
  \BibitemOpen
  \bibfield  {author} {\bibinfo {author} {\bibfnamefont {M.}~\bibnamefont
  {Ditsworth}},\
  }\bibfield  {title} {\bibinfo {title} {\texttt{github.com/markditsworth/ModularityStudy}} } {(\bibinfo {year}
  {2019})}\BibitemShut {NoStop}%
%\bibitem [{Note1()}]{Note1}%
%  \BibitemOpen
%  \bibinfo {note} {M. Ditsworth. \protect \texttt {\protect \href
%  {https://github.com/markditsworth/ModularityStudy}{github.com/markditsworth/Modularity\\Study}}
%  (2019)}\BibitemShut {NoStop}%
\bibitem [{\citenamefont {Bonacich}\ and\ \citenamefont {Lloyd}(2001)}]{EVC}%
  \BibitemOpen
  \bibfield  {author} {\bibinfo {author} {\bibfnamefont {P.}~\bibnamefont
  {Bonacich}}and\ \bibinfo {author} {\bibfnamefont {P.}~\bibnamefont {Lloyd}},\
  }\bibfield  {title} {\bibinfo {title} {Eigenvector-like measures of
  centrality for asymmetric relations},\ }\href@noop {} {\bibfield  {journal}
  {\bibinfo  {journal} {Social Networks}\ } (\bibinfo {year}
  {2001})}\BibitemShut {NoStop}%
\bibitem [{\citenamefont {Katz}(1953)}]{katz}%
  \BibitemOpen
  \bibfield  {author} {\bibinfo {author} {\bibfnamefont {L.}~\bibnamefont
  {Katz}},\ }\bibfield  {title} {\bibinfo {title} {A new status index derived
  from sociometric analysis},\ }\href@noop {} {\bibfield  {journal} {\bibinfo
  {journal} {Psychometrika}\ }\textbf {\bibinfo {volume} {18}} (\bibinfo {year}
  {1953})}\BibitemShut {NoStop}%
\bibitem [{\citenamefont {Sharkey}(2019)}]{localization}%
  \BibitemOpen
  \bibfield  {author} {\bibinfo {author} {\bibfnamefont {K.~J.}\ \bibnamefont
  {Sharkey}},\ }\bibfield  {title} {\bibinfo {title} {Localization of
  eigenvector centrality in networks with a cut vertex},\ }\href@noop {}
  {\bibfield  {journal} {\bibinfo  {journal} {Physical Review E}\ }\textbf
  {\bibinfo {volume} {99}} (\bibinfo {year} {2019})}\BibitemShut {NoStop}%
\bibitem [{\citenamefont {Arthur}\ and\ \citenamefont
  {Vassilvitskii}(2007)}]{kmeans}%
  \BibitemOpen
  \bibfield  {author} {\bibinfo {author} {\bibfnamefont {D.}~\bibnamefont
  {Arthur}}and\ \bibinfo {author} {\bibfnamefont {S.}~\bibnamefont
  {Vassilvitskii}},\ }\bibfield  {title} {\bibinfo {title} {K-means++: the
  advantages of careful seeding},\ }in\ \href@noop {} {\emph {\bibinfo
  {booktitle} {Proceedings of the eighteenth annual ACM-SIAM symposium on
  Discrete algorithms}}}\ (\bibinfo {year} {2007})\BibitemShut {NoStop}%
\bibitem [{\citenamefont {Ester}\ \emph {et~al.}(1996)\citenamefont {Ester},
  \citenamefont {Kriegel}, \citenamefont {Sander},\ and\ \citenamefont
  {Xu}}]{dbscan}%
  \BibitemOpen
  \bibfield  {author} {\bibinfo {author} {\bibfnamefont {M.}~\bibnamefont
  {Ester}}, \bibinfo {author} {\bibfnamefont {H.~P.}\ \bibnamefont {Kriegel}},
  \bibinfo {author} {\bibfnamefont {J.}~\bibnamefont {Sander}}, and\ \bibinfo
  {author} {\bibfnamefont {X.}~\bibnamefont {Xu}},\ }\bibfield  {title}
  {\bibinfo {title} {A density-based algorithm for discovering clusters in
  large spatial databases with noise},\ }in\ \href@noop {} {\emph {\bibinfo
  {booktitle} {Proceedings of the 2nd International Conference on Knowledge
  Discovery and Data Mining}}}\ (\bibinfo {year} {1996})\BibitemShut {NoStop}%
\bibitem [{\citenamefont {Voorhees}(1989)}]{agglom}%
  \BibitemOpen
  \bibfield  {author} {\bibinfo {author} {\bibfnamefont {E.~M.}\ \bibnamefont
  {Voorhees}},\ }\href@noop {} {\emph {\bibinfo {title} {Implementing
  Agglomerative Hierarchical Clustering Algorithms for Use in Document
  Retrieval}}}\ (\bibinfo  {publisher} {Cornell University Department of
  Computer Science},\ \bibinfo {year} {1989})\BibitemShut {NoStop}%
\bibitem [{\citenamefont {Radon}(1917)}]{radon}%
  \BibitemOpen
  \bibfield  {author} {\bibinfo {author} {\bibfnamefont {J.}~\bibnamefont
  {Radon}},\ }\bibfield  {title} {\bibinfo {title} {Uber die bestimmung von
  funktionen durch ihre integralwerte langs gewisser mannigfaltigkeiten},\
  }\href@noop {} {\bibfield  {journal} {\bibinfo  {journal} {Journal of
  Mathematical Physics}\ }\textbf {\bibinfo {volume} {69}},\ \bibinfo {pages}
  {262} (\bibinfo {year} {1917})}\BibitemShut {NoStop}%
\bibitem [{\citenamefont {Erd\H{o}s}\ and\ \citenamefont {R\'enyi}(1960)}]{ER}%
  \BibitemOpen
  \bibfield  {author} {\bibinfo {author} {\bibfnamefont {P.}~\bibnamefont
  {Erd\H{o}s}}and\ \bibinfo {author} {\bibfnamefont {A.}~\bibnamefont
  {R\'enyi}},\ }\bibfield  {title} {\bibinfo {title} {On the evolution of
  random graphs},\ }\href@noop {} {\bibfield  {journal} {\bibinfo  {journal}
  {Publications of the Mathematical Institute of the Hungarian Academy of
  Sciences}\ }\textbf {\bibinfo {volume} {5}} (\bibinfo {year}
  {1960})}\BibitemShut {NoStop}%
\bibitem [{\citenamefont {Barab\'asi}\ and\ \citenamefont {Albert}(1999)}]{BA}%
  \BibitemOpen
  \bibfield  {author} {\bibinfo {author} {\bibfnamefont {A.~L.}\ \bibnamefont
  {Barab\'asi}}and\ \bibinfo {author} {\bibfnamefont {R.}~\bibnamefont
  {Albert}},\ }\bibfield  {title} {\bibinfo {title} {Emergence of scaling in
  random networks},\ }\href@noop {} {\bibfield  {journal} {\bibinfo  {journal}
  {Science}\ }\textbf {\bibinfo {volume} {286}} (\bibinfo {year}
  {1999})}\BibitemShut {NoStop}%
\bibitem [{\citenamefont {Andris}\ \emph {et~al.}(2015)\citenamefont {Andris},
  \citenamefont {Lee}, \citenamefont {Hamilton}, \citenamefont {Martino},
  \citenamefont {Gunning},\ and\ \citenamefont {Selden}}]{congress}%
  \BibitemOpen
  \bibfield  {author} {\bibinfo {author} {\bibfnamefont {C.}~\bibnamefont
  {Andris}}, \bibinfo {author} {\bibfnamefont {D.}~\bibnamefont {Lee}},
  \bibinfo {author} {\bibfnamefont {M.~J.}\ \bibnamefont {Hamilton}}, \bibinfo
  {author} {\bibfnamefont {M.}~\bibnamefont {Martino}}, \bibinfo {author}
  {\bibfnamefont {C.~E.}\ \bibnamefont {Gunning}}, and\ \bibinfo {author}
  {\bibfnamefont {J.~A.}\ \bibnamefont {Selden}},\ }\bibfield  {title}
  {\bibinfo {title} {The rise of partisanship and super-cooperators in the u.s.
  house of representatives},\ }\href@noop {} {\bibfield  {journal} {\bibinfo
  {journal} {PLOS ONE}\ }\textbf {\bibinfo {volume} {10}} (\bibinfo {year}
  {2015})}\BibitemShut {NoStop}%
\bibitem [{\citenamefont {Yang}\ and\ \citenamefont {Leskovec}(2012)}]{DBLP}%
  \BibitemOpen
  \bibfield  {author} {\bibinfo {author} {\bibfnamefont {J.}~\bibnamefont
  {Yang}}and\ \bibinfo {author} {\bibfnamefont {J.}~\bibnamefont {Leskovec}},\
  }\bibfield  {title} {\bibinfo {title} {Defining and evaluating network
  communities based on ground-truth},\ }in\ \href@noop {} {\emph {\bibinfo
  {booktitle} {ICDM}}}\ (\bibinfo {year} {2012})\BibitemShut {NoStop}%
\bibitem [{\citenamefont {Leskovec}\ and\ \citenamefont {Krevl}(2014)}]{SNAP}%
  \BibitemOpen
  \bibfield  {author} {\bibinfo {author} {\bibfnamefont {J.}~\bibnamefont
  {Leskovec}}and\ \bibinfo {author} {\bibfnamefont {A.}~\bibnamefont {Krevl}},\
  }\href@noop {} {\bibinfo {title} {{SNAP Datasets}: {Stanford} large network
  dataset collection}},\ \bibinfo {howpublished}
  {\url{http://snap.stanford.edu/data}} (\bibinfo {year} {2014})\BibitemShut
  {NoStop}%
\bibitem [{\citenamefont {Mislove}\ \emph {et~al.}(2007)\citenamefont
  {Mislove}, \citenamefont {Marcon}, \citenamefont {Gummadi}, \citenamefont
  {Druschel},\ and\ \citenamefont {Bhattacharjee}}]{youtube}%
  \BibitemOpen
  \bibfield  {author} {\bibinfo {author} {\bibfnamefont {A.}~\bibnamefont
  {Mislove}}, \bibinfo {author} {\bibfnamefont {M.}~\bibnamefont {Marcon}},
  \bibinfo {author} {\bibfnamefont {K.~P.}\ \bibnamefont {Gummadi}}, \bibinfo
  {author} {\bibfnamefont {P.}~\bibnamefont {Druschel}}, and\ \bibinfo {author}
  {\bibfnamefont {B.}~\bibnamefont {Bhattacharjee}},\ }\bibfield  {title}
  {\bibinfo {title} {{Measurement and Analysis of Online Social Networks}},\
  }in\ \href@noop {} {\emph {\bibinfo {booktitle} {Proceedings of the 5th
  ACM/Usenix Internet Measurement Conference (IMC'07)}}}\ (\bibinfo {address}
  {San Diego, CA},\ \bibinfo {year} {2007})\BibitemShut {NoStop}%
\bibitem [{\citenamefont {McAuley}\ \emph {et~al.}(2015)\citenamefont
  {McAuley}, \citenamefont {Targett}, \citenamefont {Shi},\ and\ \citenamefont
  {van~den Hengle}}]{amazon}%
  \BibitemOpen
  \bibfield  {author} {\bibinfo {author} {\bibfnamefont {J.}~\bibnamefont
  {McAuley}}, \bibinfo {author} {\bibfnamefont {C.}~\bibnamefont {Targett}},
  \bibinfo {author} {\bibfnamefont {J.}~\bibnamefont {Shi}}, and\ \bibinfo
  {author} {\bibfnamefont {A.}~\bibnamefont {van~den Hengle}},\ }\bibfield
  {title} {\bibinfo {title} {Image-based recommendations on styles and
  substitutes},\ }in\ \href@noop {} {\emph {\bibinfo {booktitle} {SIGIR}}}\
  (\bibinfo {year} {2015})\BibitemShut {NoStop}%
\bibitem [{\citenamefont {Pujol}\ \emph {et~al.}(2010)\citenamefont {Pujol},
  \citenamefont {Erramilli}, \citenamefont {Siganos}, \citenamefont {Yang},
  \citenamefont {Laoutaris}, \citenamefont {Chhabra},\ and\ \citenamefont
  {Rodriguez}}]{louvainDist}%
  \BibitemOpen
  \bibfield  {author} {\bibinfo {author} {\bibfnamefont {J.~M.}\ \bibnamefont
  {Pujol}}, \bibinfo {author} {\bibfnamefont {V.}~\bibnamefont {Erramilli}},
  \bibinfo {author} {\bibfnamefont {G.}~\bibnamefont {Siganos}}, \bibinfo
  {author} {\bibfnamefont {X.}~\bibnamefont {Yang}}, \bibinfo {author}
  {\bibfnamefont {N.}~\bibnamefont {Laoutaris}}, \bibinfo {author}
  {\bibfnamefont {P.}~\bibnamefont {Chhabra}}, and\ \bibinfo {author}
  {\bibfnamefont {P.}~\bibnamefont {Rodriguez}},\ }\bibfield  {title} {\bibinfo
  {title} {The little engine(s) that could: scaling online social networks},\
  }in\ \href@noop {} {\emph {\bibinfo {booktitle} {SIGCOMM '10}}}\ (\bibinfo
  {year} {2010})\BibitemShut {NoStop}%
\end{thebibliography}
\end{document}